\documentclass[preprint2]{aastex}

\shorttitle{On the bar pattern speed of NGC 3367}
\shortauthors{Gabbasov et al.}

\begin{document}

\title{On the bar pattern speed determination of NGC 3367}
%\title{Pattern speed determination of NGC 3367 using H${\alpha}$ emission.}

\author{R. F. Gabbasov, P. Repetto, M. Rosado}
\affil{Instituto de Astronom\'{i}a, Universidad Nacional
Aut\'onoma de Mexico (UNAM), A.P. 70-264,04510, M\'exico D.F.}
\email{gabbasov@astroscu.unam.mx}

\begin{abstract}

An important dynamic parameter of barred galaxies is the bar
pattern speed, $\Omega_P$. Among several methods that are used for
the determination of $\Omega_P$ the Tremaine-Weinberg method has
the advantage of model independency and accuracy. In this work we
apply the method to a simulated bar including gas dynamics and
study the effect of 2D spectroscopy data quality on robustness of
the method. We added a white noise and a Gaussian random field to
the data and measured the corresponding errors in $\Omega_P$. We
found that a signal to noise ratio in surface density $\sim5$
introduces errors of $\sim 20\%$ for the Gaussian noise, while for
the white noise the corresponding errors reach $\sim 50\%$. At
the same time the velocity field is less sensitive to
contamination. On the basis of the performed study we applied the
method to the NGC 3367 spiral galaxy using H${\alpha}$ Fabry-Perot
interferometry data. We found $\Omega_P=43\pm 6$ km s$^{-1}$ kpc$^{-1}$ 
for this galaxy.
\end{abstract}


\section{Introduction}

The evolution of disk galaxies is strongly influenced by the main
structures within their disks, notably spirals and bars.
Bars are a fundamental component of mass
distribution in spiral galaxies because they may contain a large
fraction of the disk. Bar kinematics is different from the one of
the axisymmetric disk, which makes the study of bar parameters
important, especially the bar pattern speed ($\Omega_P$). Bar
properties are tightly related to the mass distribution of the
host galaxy. The shape and structure of the bar, such as vertical
bending (buckling), is believed to depend on the stage of
evolution \citep{Athanassoula2003}. Moreover, $\Omega_P$ is one of
the fundamental parameters in driving the evolution of bars.
\citet{Noguchi1987} finds from numerical simulations that
tidally-induced bars rotate slowly in comparison with spontaneous
bars. Thus, bar pattern speed may help to discriminate between
spontaneous bars and tidal bars.

One of the first methods applied to bar pattern speed
determination is based on the identification of theoretically
predicted resonances (Lindblad resonances, corotation), using the
rotation curve to extract the periodic motion of stars and gas
\citep{Tully1974}. \citet{Contopoulos1980} demonstrated with orbit
calculations that the corotation radius should be located at the
end of a self-consistent bar. However, it was shown from a survey
of early-type galaxies, that indeed the bar ends rather between
the inner $4:1$ resonance and the corotation
\citep{Elmegreen1996}.

An alternative method consists of matching numerical models to the
observed data \citep{Rautiainen2005}. Two other methods are based
on the analysis of spiral density waves. \citet{Elmegreen1992}
apply a computer algorithm to extract various types of symmetries
from galaxy images, while \citet{Canzian1993} points out the difference in
the global appearance of the residual velocity field of a spiral
galaxy inside and outside the corotation radius. \citet{Tremaine1984}
formulated a method to measure the bar pattern speed that is
independent from any spiral density wave theory. The principal
assumption of the method is that the surface density of the tracer
of the gravitational potential (e.g., old stars) satisfies the
continuity equation, i.e., there is no significant destruction or
creation of the tracer over a dynamical time. This method allows
the measurement of the bar pattern speed with two observational
quantities: the surface brightness of the tracer and the velocity
of the tracer along the line of sight. So far the method was
successfully applied to some twenty galaxies. \citet{Kent1987} and
\citet{Merrifield1995} apply this method to the stellar bar
component of the SB0 galaxy NGC 936. \citet{Gerssen1999} apply the
Tremaine-Weinberg (TW) method to the stellar component of the
galaxy NGC 4596. The same authors use the method for the galaxies
NGC 271, NGC1358, ESO 281-31 and NGC 3992 \citep{Gerssen1999}.
\citet{Aguerri2003} applied the method for the galaxies ESO
139-G009, IC 874, NGC 1308, NGC 1440 and NGC 3412.
\citet{Debattista2001} use the TW method for NGC 7079 and for NGC
1023 \citep{Debattista2002}.

Other authors derived the bar pattern speed applying the TW method
to a gaseous tracer. \citet{Westpfahl1998} finds the bar pattern
speed for M81 employing HI as a tracer. Using CO observations,
\citet{Zimmer2004} employ the method for the galaxies M51, M83 and
NGC 6946. \citet{Rand2004} consider the galaxies NGC 1068, NGC
3627, NGC 4321 (M100), NGC 4414, NGC 4736 and NGC 4826, and
measure the bar pattern speed with CO as a tracer. The application
of this method to the gaseous phase is more delicate because of
the assumptions of the method itself. In general, the ionized gas,
the CO and HI, do not satisfy the continuity equation over an
orbital period and do not trace the gravitational potential.
However, assuming that the { gas is continuously distributed along
the pattern}, and that the luminosity-weighted mean is a valuable
indicator of the mean mass distribution and, neglecting the
internal kinematics of H II regions, some authors successfully
applied the TW method to the ionized gaseous phase.
\citet{Hernandez2004} employ ionized hydrogen 2D velocity fields
of four barred galaxies NGC 4321, NGC 3359, NGC 6946, NGC 2903 and
also of M51 to measure the pattern speed. The same authors build
numerical simulations to investigate the possibility of using the
gaseous component and present an application for NGC 4321 galaxy
\citep{Hernandez2005}.  \citet{Emsellem2006} have successfully
applied TW method for bar pattern speed determination of NGC 1068
using Fabry-Perot H$\alpha$ map, and also reproduce with Nbody+SPH
models a number of its observed properties. \citet{Fathi2007}
also with two dimensional ionized hydrogen kinematics determine
the bar pattern speed of NGC 6946. \citet{Beckman2007} apply the
method to a sample of nine galaxies (NGC 3049, NGC 4294, NGC4519,
NGC 5371, NGC 5921, NGC 5964, NGC 6946, NGC 7479, NGC 7741) using
ionized hydrogen 2D kinematic data.

In this article we employ the TW method to study
the errors in the parameters that may affect the pattern speed
determination. We test to which parameters the method is
sensitive with numerical simulations and apply this method to the
barred galaxy NGC 3367.

%%%
%%%%%%%%%%%%%%%%%%%
%%%

\section{Numerical test of the TW method} \label{twtest}
The Tremaine Weinberg method uses two observable quantities: the
luminosity weighted velocity and luminosity weighted density
determined along a thin strip (aperture) parallel to the major
axis of the disk \citep{Tremaine1984}. If a galaxy is centered at
the cartesian coordinates such that the major and minor axes are
aligned parallel to the $x$ and $y$ axes, respectively, then the
ratio of intensity-weighted velocity and intensity-weighted
position gives the angular velocity:
\begin{equation}
\Omega_P\sin
i=\frac{\langle V(x)\rangle-V_{sys}}{\langle x\rangle-x_0}.
%\frac{\int^{\infty}_{-\infty}\Sigma(x)[V_{los}(x)-V_{sys}]dx}
%{\int^{\infty}_{-\infty}\Sigma(x)[x-x_0]dx}.
\label{eq1}
\end{equation}
Here $V_{sys}$ - is the systemic velocity, and $x_0$ is the
position of the kinematic center of the disk along $x$ axis. The
$\Omega_P$ is also corrected due to inclination $i$ ($i=0$ corresponds to
face-on disk). This formulation of
\citet{Merrifield1995} allows a more accurate evaluation of
$\Omega_P$ because the errors in the dynamical center and systemic
velocity determination are reduced. Thus, estimating the above
expression for several apertures, one may plot $\left<V\right>$
vs. $\left<x\right>$ and obtain the averaged value of $\Omega_P$.

One of the conditions required by the TW method is that the tracer
should satisfy the continuity equation. The old stars in SB0
galaxies, for example, survive long enough to trace the spiral
pattern potential. The main concern regarding the validity of
application of the TW method to gaseous tracer is that the ISM,
composed mostly of molecular and atomic hydrogen, is not able to
trace the potential for a long period of time due to short
timescale processes such as phase transitions, cooling, etc.
Nevertheless, it was successfully applied to CO emission lines
\citep{Rand2004,Zimmer2004} and to the H${\alpha}$ emission line
\citep{Hernandez2005,Emsellem2006,Fathi2007}. As was shown by
\citet{Westpfahl1998} addition or substraction of the tracer
material has no effect on calculation of the instantaneous pattern
speed, and thus, it is expected that processes of star formation
and feedback may be neglected. \citet{Hernandez2005} have
demonstrated by hydrodynamical simulations that the application of
the method to ionized gaseous phase is reliable under some
assumptions. The problem arises rather in the quality of the image
and the velocity field as affected by regions of star formations,
dust obscuration, and local gas motions. An argument in favor of
applicability of the method is that the H${\alpha}$ image also
contains the stellar continuum emission. This should alleviate the
problem of a patchy monochromatic image, making it smoother. All
these arguments are still have to be carefully verified, and here
we assume as a working hypothesis that application of the TW
method to the H${\alpha}$ kinematic data is valid.

As it was shown by \citet{Debattista2003} the method is very
sensitive to errors in the determination of the position angle
(PA) of the major axis. On the other hand, as shown by
\citet{Rand2004}, the molecular gaseous component may produce a
non-zero pattern speed even in the absence of any clear wave
pattern due to clumpiness. The latter point requires that the
determination of $\Omega_P$ should be done with much caution. In
particular, \citet{Hernandez2005} have shown by means of
hydrodynamical simulations that the regions of shocks and the
zones outside the bar should be avoided. The effects of the
inclination, bar orientation, angular resolution, and the
uncertainty were already investigated in numerical models by
\citet{Rand2004}.

\subsection{Numerical models}
In order to test the influence of errors in intensity and 2D
velocity fields on the determination of the pattern speed with the
TW method we performed numerical simulations of a bar formed in a
 hydrodynamic (Nbody+SPH) and collisionless (Nbody) disk galaxy.

\begin{table}
\begin{center}
\caption{Galaxy model parameters.\label{tbl-1}}
\begin{tabular}{lccl}
\tableline \tableline
 Parameter & Gas & Disk & Halo \\
\tableline
 $M$ ($10^{10}{\mbox M_{\odot}}$) & $0.4$ & $3.96$ & $22$ \\
 r$_0$ (kpc) & $3.3$ & $3.3$ & $6.6$ \\
 $N$ ($10^{6}$)& $0.2$ & $0.8$ & $2.0$ \\
 $\varepsilon$ (pc) & $80$ & $80$ & $160$ \\
\tableline
\end{tabular}
\tablecomments{Here M is the mass of the component, r$_0$ - the
scale radius, $N$ - the number of particles, and $\varepsilon$ is
the softening length.}
\end{center}
\end{table}

For this purpose we prepared a bar unstable disk galaxy consisting
of total $3\times 10^6$ particles sampling an exponential
stellar and gas disks, and a Hernquist halo
\citep{Hernquist1990}. The vertical structure of the disk is
described by isothermal sheets with a constant scale height of
$280$ pc { for stars and $80$ pc for gas. Gas dynamics obeys the
isothermal equation of state with a temperature T$=10^4$ K. The
processes of star formation and feedback were not included.} The
model was constructed using a technique similar to that described
by \citet{Hernquist1993} and the numerical parameters were chosen
according to \citet{Gabbasov2006}. The galaxy model parameters
are summarized in the Table~\ref{tbl-1}. The simulation was
performed with GADGET 2 code on HP CP 4000 cluster (KanBalam) at
DGSCA-UNAM.

\begin{figure}[!htp]
\plotone{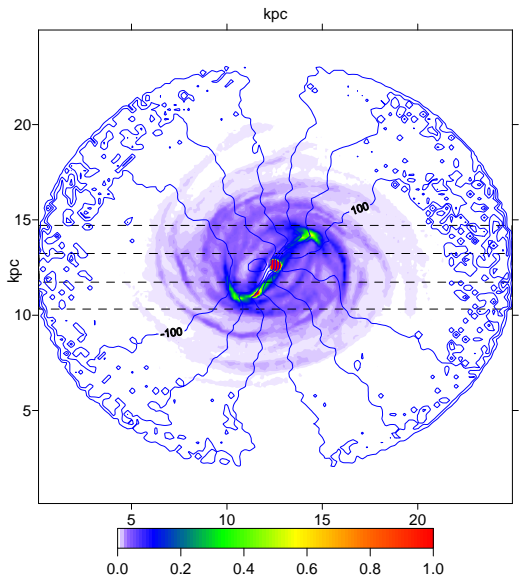} \figcaption{Projected
surface density of the gas, with the corresponding velocity
contours oveplotted. The surface density is normalized by the
maximum value. The contours are separated by $40$ km s$^{-1}$ and
dashed lines show some of the slit positions (one of ten).
\label{fig1}}
\end{figure}

We estimate the bar pattern speed as follows. At each snapshot we
determine the orientation of the principal axes of the inertia
tensor of the bar. Then, we draw the bar position angle as a
function of time, and the $\Omega_P$ is obtained by numerical
differentiation, $\Omega_P=d\phi/dt$. The bar appears at a time
$t\sim 2$ Gyrs and at $t=3$ Gyrs reaches its maximum length of
$\sim7$ kpc. For $t=4$ Gyrs most of the gas is transferred to the
center of the disk. During this period, the angular velocity of
the bar remains roughly constant, $41$ km s$^{-1}$. Further
evolution leads to the depletion of the gas particles along the
bar, and after about $5$ Gyrs, the gas is rather located in the
center and in the spiral arms. For our analysis we took the
snapshot at $t=3.75$ Gyrs when the bar is oriented $\sim 40\degr$
from the vertical $y$ axis. We added some systemic velocity to
the velocity field and inclined the disk by $i=30\degr$, such that
the disk major axis is aligned with the $x$ axis. We compute the
projected on the sky plane 2D surface density and the velocity
fields on uniform cartesian grids of $204 \times 204$, see
Fig.~\ref{fig1}. We apply the TW method to simulated stellar 
and gas bars and compared $\Omega_P$ with the pattern speed
obtained from the simulation. A difference of less than $1.0$ km
s$^{-1}$ was found for the stellar bar. The gas pattern speed
was overestimated by $\sim 7$ km s$^{-1}$, because of the surface
density weighting errors. We also applied the method to the
galaxy at $t=1$ Gyrs (early stage of evolution without a bar) and
obtained a zero slope as expected.

\begin{figure}[!htp]
\plotone{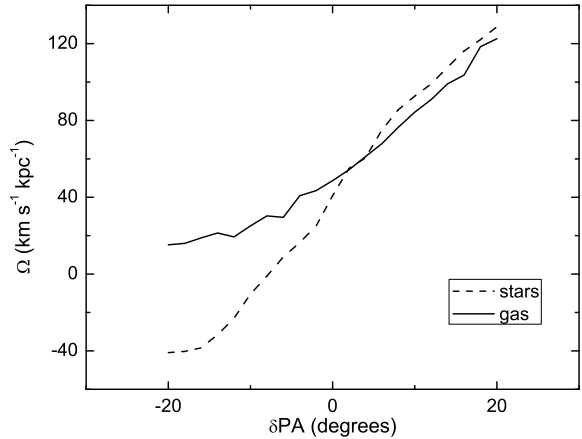} \figcaption{Dependence of
$\Omega_P$ on the PA variation. \label{fig2}}
\end{figure}

\subsection{PA variation}
First, we tested the sensitivity of the bar pattern speed to
variations of the position of the major axis of the simulated disk
galaxy, which could be interpreted as errors in the determination
of PA. Here, the variation in the position angle $\delta$PA, is
the angle between the disk major axis and the $x$ axis of the
fixed reference frame. The resulting $\Omega_P$ is presented in
Fig.~\ref{fig2}. For the considered range of $\delta$PA  and bar
orientation the $\Omega_P$ changes almost linearly. As the bar
tends to become aligned with the $y$ axis the pattern speed
reduces. From this plot one may observe that an uncertainty of
$\delta$PA $\pm 5\degr$ gives rise to an error in $\Omega_P$ of
$\pm 15$ km s$^{-1}$kpc$^{-1}$ for gas bar.

\begin{figure*}[!htp]
\epsscale{2.0}\plottwo{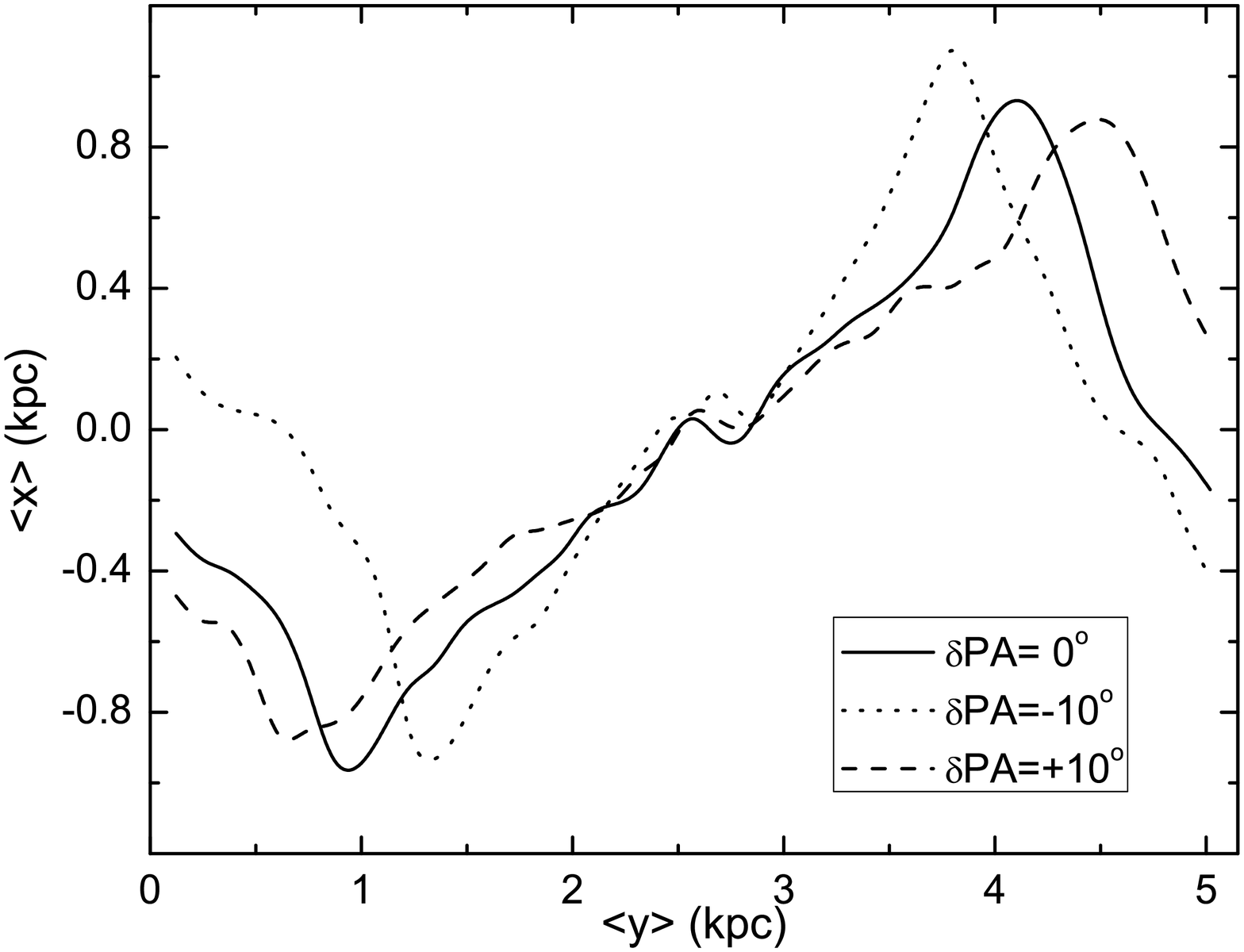}{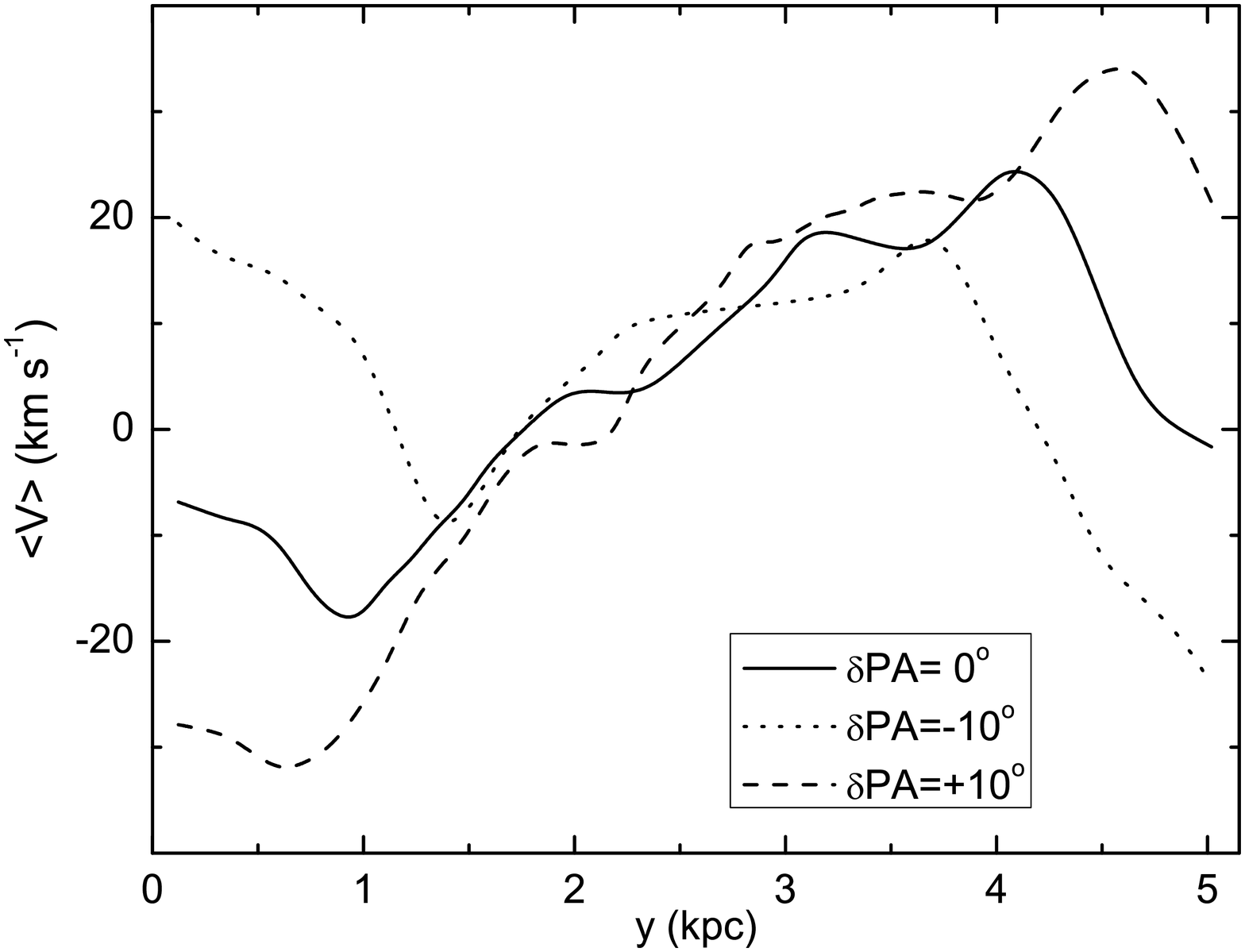} \figcaption{Dependence
of $\langle X\rangle$ and $\langle V\rangle$ on PA
variation.\label{fig3}}
\end{figure*}

\begin{figure*}[!htp]
\epsscale{2.0}\plottwo{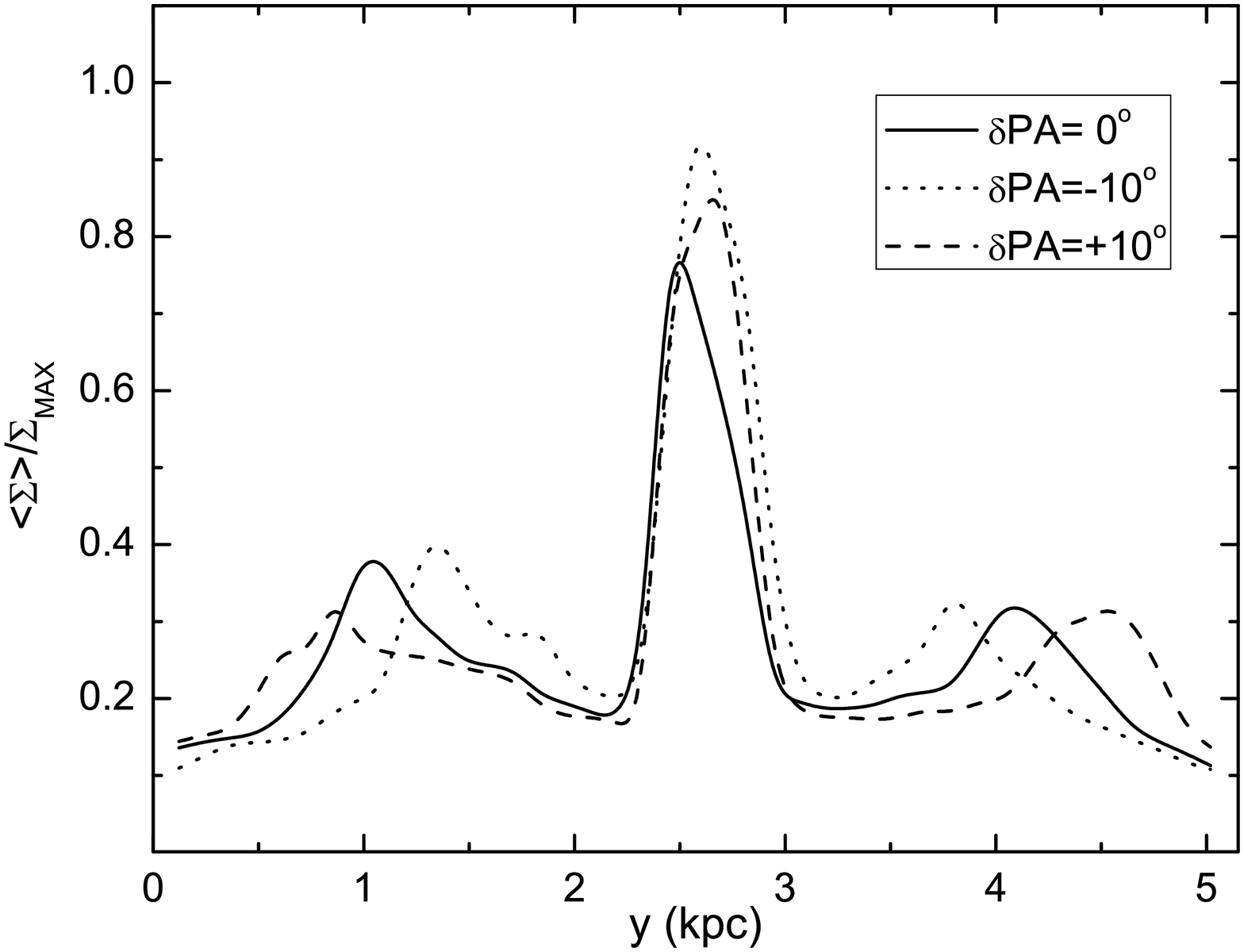}{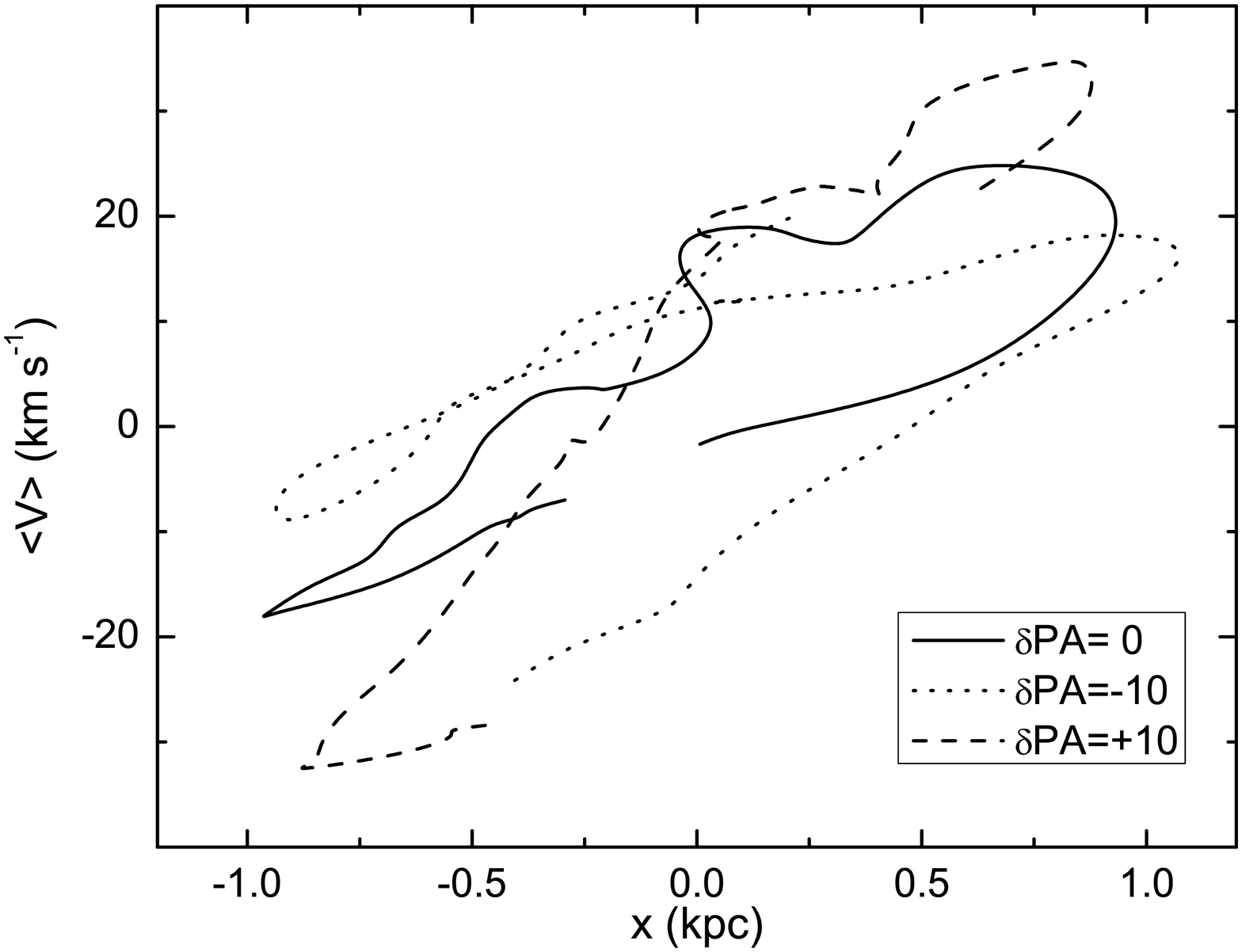} \figcaption{Dependence
of $\langle\Sigma\rangle$ and the slope $\langle V\rangle$ vs.
$\langle x\rangle$ on PA variation.\label{fig4}}
\end{figure*}

As an example, in Fig.~\ref{fig3} we plot the $\left<x\right>$,
$\left<V\right>$, and { in Fig.~\ref{fig4} the normalized} mean
surface density $\left<\Sigma\right>/\Sigma_{MAX}$ vs. $y$ for
$40$ apertures { of width $0.12$ kpc} about the kinematic center
located at $y_0=2.45$ kpc. Also in Fig.~\ref{fig4} shown is the
resulting $\left<V\right>$ to $\left<x\right>$ ratio. In order
to avoid a discontinuity in center of these plots we have excluded
from the analysis the central four pixels with very high gas
density. The plots are shown for $\delta$PA$=0,-10\degr,+10\degr$
(solid, dotted, and dashed line, respectively).

As seen from Fig.~\ref{fig3}, the weighted average velocity is the
most sensitive quantity in the plot, and the difference comes
mainly from the ends of the bar, while at the center the slope
varies slowly for all three cases. At the same time the slope of
$\left<x\right>$ changes in opposite direction than
$\left<V\right>$, affecting strongly the $\Omega_P$ determined. {
The decaying intensity and velocity profiles in the first and the
last kiloparsec are due to inclusion of the apertures containing
spiral arms at the ends of the bar.} This is observed as a
characteristic $\mathcal{Z}$-shape, or a loop-shape, in the
$\left<V\right>$ vs. $\left<x\right>$ plot in Fig.~\ref{fig4},
instead of a straight line. The weighted averaged intensity also
changes the position and the maximum of its shoulders, being
nearer to the center and higher for closer alignment of the bar
with the $y$ axis. Unfortunately, the restriction in $y$ does not
reduce significantly the errors produced by variation of PA.

\subsection{Disk thickness}
\begin{figure*}[!htp]
\epsscale{1.6}\plotone{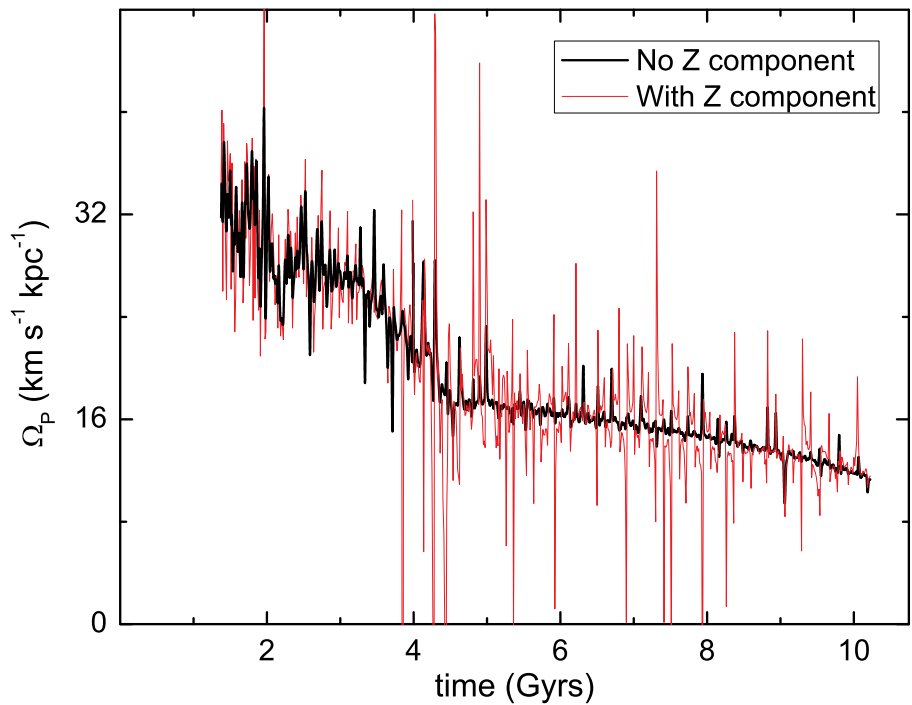} \figcaption{Pattern speed of a
collisioness bar determined with TW method, including and
excluding the Z component (red thin and black thick curve,
respectively).\label{fig5}}
\end{figure*}

According to the original formulation of the method, it is valid
only for an infinitely thin disk. However, galactic disks
generally have a certain thickness and line of sight velocities
also contain the projected component of velocity normal to the
disk ($Z$ component, $v_z\cos i$). This component becomes
important as the inclination approaches $0$ degrees, i.e., the
galaxy becomes face on. For a disk symmetric in the $Z$ direction
the vertical motions cancel out each other, and the net
contribution is zero. However,the presence of an asymmetry in
vertical structure (such as bar buckling) will also introduce
asymmetry in the velocity field. We have verified this argument by
including the projected $Z$ component of velocity in a simulated
bar velocity field. For our Nbody+SPH model, evolved for $5$
Gyrs, we found no significant difference in $\Omega_P$ for both
gas and stellar bars. We also prepared and ran a pure
collisionless model, incrementing the stellar disk mass by the
mass of removed gas disk. We traced the evolution of the bar up to
$10$ Gyrs, and quantify the bar buckling as a change of r.m.s. of
the vertical velocity dispersion of disk particles, $\sigma_z$.
Due to heating of the stellar disk the vertical dispersions grow
linearly until the bar begins to buckle at $t\approx 4.5$ Gyrs.
After the buckling, the r.m.s. dispersions are rapidly increased
by $\sim 10\%$ and this is reflected in a reduction of the pattern
speed slowdown rate, as shown in Fig.~\ref{fig5}. The
determination of the pattern speed becomes more robust as the
spiral arms cease, and the curve that does not include the Z
component is much smoother, except for some spikes. In contrast,
inclusion of the Z component clearly affects the instantaneous
$\Omega_P$ determination due to buckling, but the errors are quite
small (a few km s$^{-1}$). Thus, for old stellar bars, where
buckling is strong and the velocity dispersions are high, the
vertical motions could make an important contribution to the
observed velocity field introducing errors in the $\Omega_P$
determination by the TW method.
\begin{figure}[!htp]
\epsscale{2.0}\plottwo{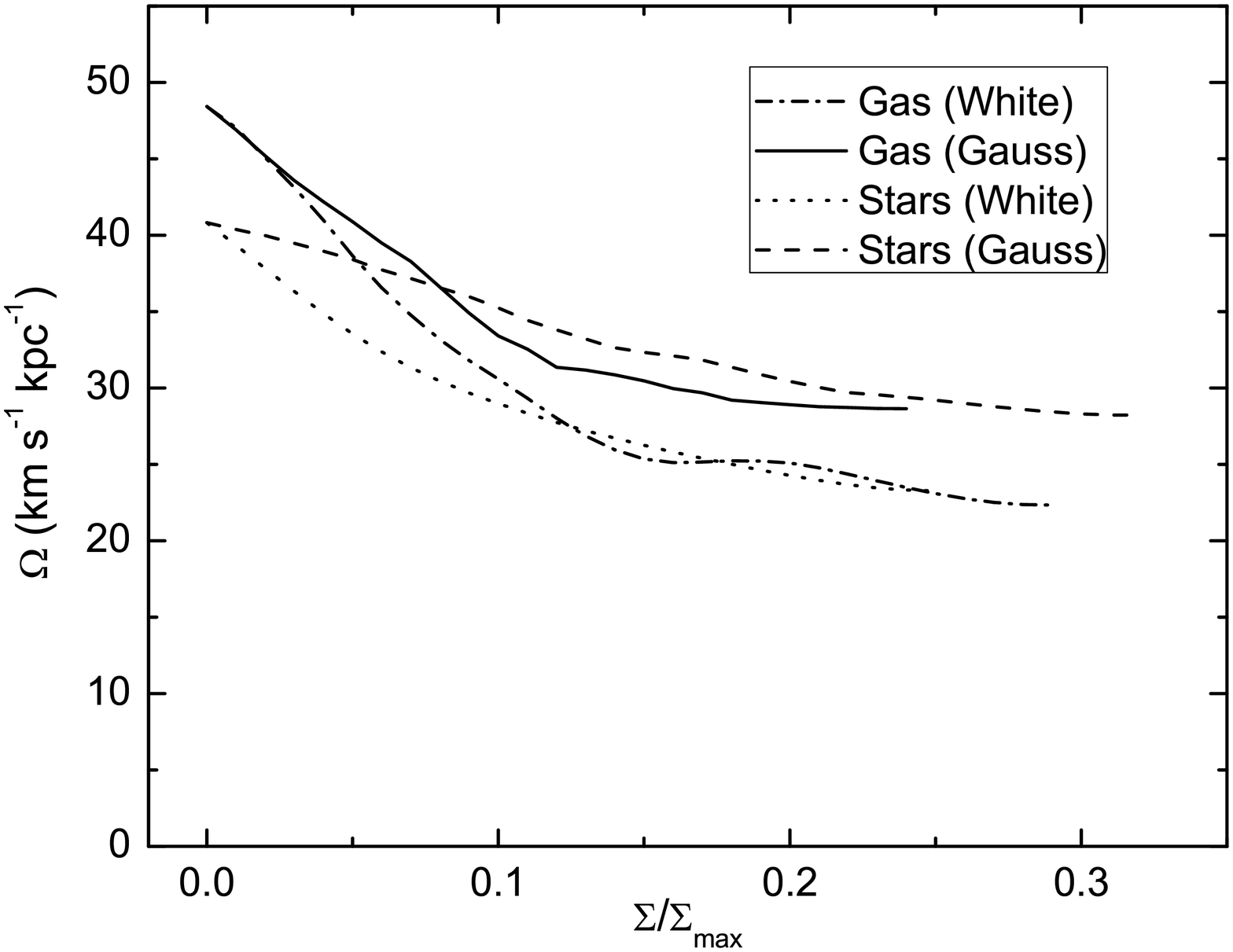}{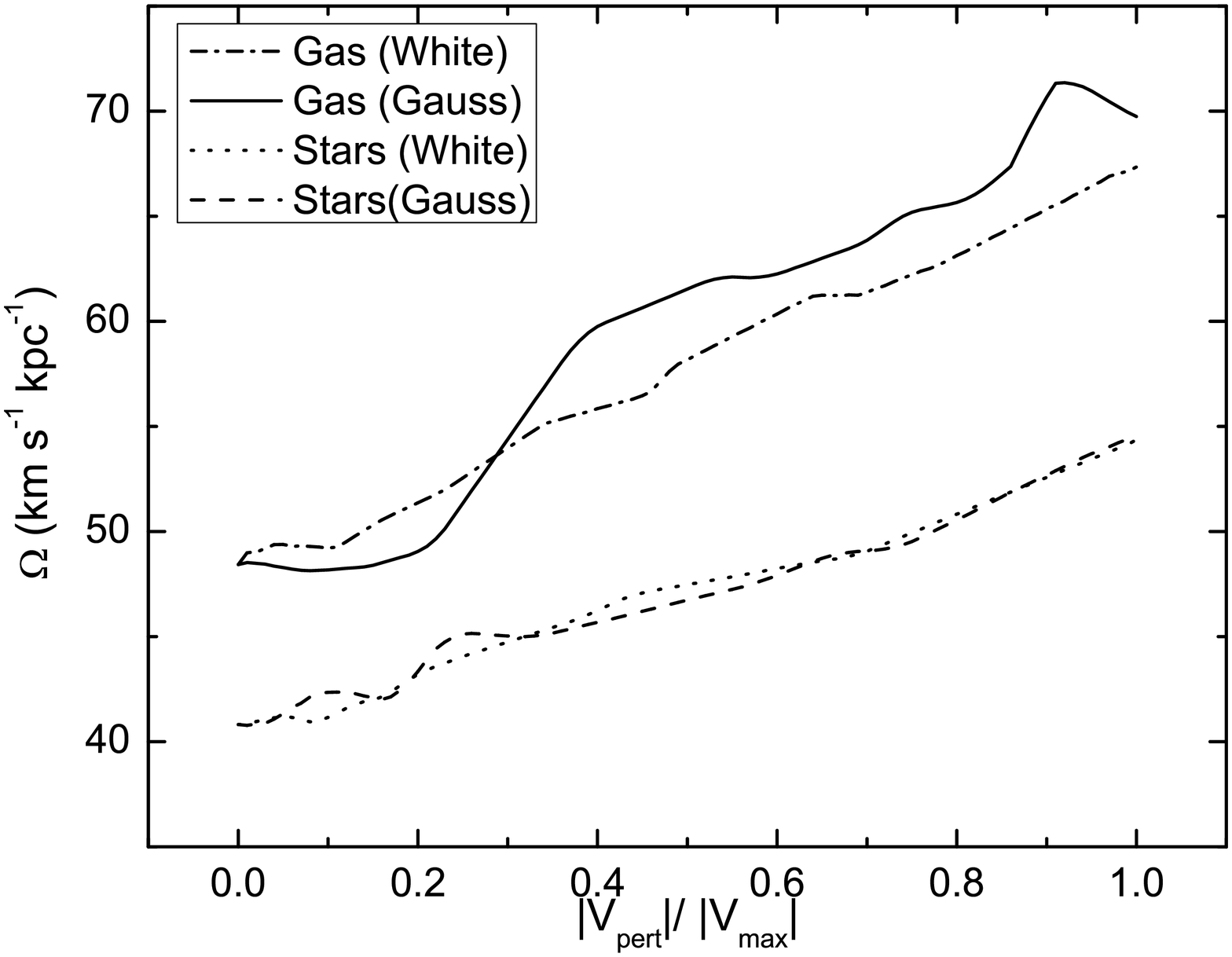} \figcaption{Variation of
$\Omega_P$ with the degree of contamination of intensity (top) and
velocity (bottom).\label{fig6}}
\end{figure}

\subsection{Data quality}
\begin{figure}[!htp]
\epsscale{1.0}\plotone{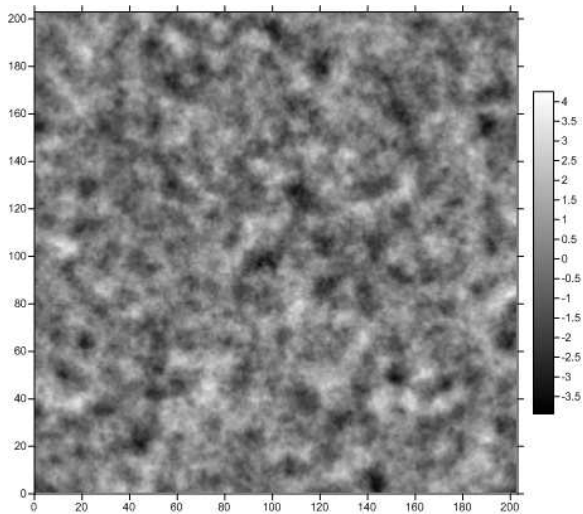} \figcaption{Random Gauss field 
used for data quality study.\label{gauss1}}
\end{figure}

Next, we checked the robustness of the TW method to the
introduction of an artificial noise to the surface density and
velocity fields. Such noise is naturally present in the observed
data. First, we add a white noise background to our simulated
surface density field to represent the emission inhomogeneities.
The white noise background was created by assigning to each pixel
a random value within ($0,\Sigma_{\rm pert}$) in case of intensity
field and $(-V_{\rm pert},+V_{\rm pert}$) in case of velocity
field. Here, $\Sigma_{\rm pert}$ and $V_{\rm pert}$ are the
maximum values of perturbation of the surface density and velocity
field, respectively. The results are shown in Fig.~\ref{fig6}. As
we increment the amplitude of the perturbation relative to the
corresponding maximum value of the surface density ($\Sigma_{\rm
pert}/\Sigma_{\rm max}$) or the velocity field ($|V_{\rm
pert}|/|V_{\rm max}|$), the errors become dominant. Indeed, for
the ratio $\Sigma_{\rm pert}/\Sigma_{\rm max}>0.3$ (signal to
noise ratio $<5$) for white noise, the determination of the
pattern speed is not possible anymore because a clear slope cannot
be established. As a next step, we test the sensitivity of the
method to the Gaussian noise. Besides the observational noise it
simulates local gas inhomogeneities in the disk. We create a
random Gaussian field with spherical ($\theta=10$) correlation
function and also change its amplitude relative to the maximum
value of the signal (see Fig.~\ref{gauss1}). 
For this purpose, we used the software
developed by \citet{kozintsev2000}. When added to the intensity
field, the signal is also underestimated, although not so strongly
as in the case of white noise. Concerning the velocity field
contamination, one can note that the determination of the
$\Omega_P$ is possible, even reaching the ratio $|V_{\rm
pert}|/|V_{\rm max}|\sim 1$. However, in the case of the
Gaussian random field the errors influence stronger the resulting
$\langle V\rangle$ vs. $\langle x\rangle$ slope for the gas bar.
Note, that the pattern speed is either underestimated or
overestimated for different bar positions and, in case of the
Gaussian noise, for different realizations of the field.

\begin{figure}[!htp]
\epsscale{1.0}\plotone{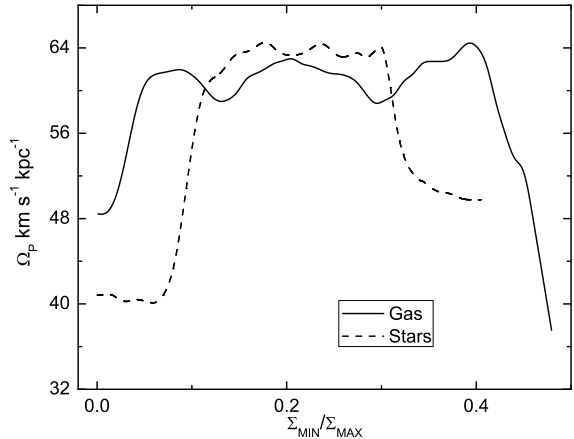} \figcaption{Variation of $\Omega_P$
with the degree of clipping of the surface density.\label{fig7}}
\end{figure}

In addition, we investigate the effect of a clipping procedure
(i.e., imposing the inferior limit on the surface density) and
finding that the effect on the resulting pattern speed has rather
non-linear trend as can be seen in Fig.~\ref{fig7}.  The surface
densities of both stellar and gas components were normalized by
their respective maximum values. The most sensitive part of the
curve is due to clipping of up to a few percent of the maximum
surface density, which corresponds to the density of the faint
spiral arms and the disk that leads to an overestimation by more
than $50\%$ in the pattern speed (cf. surface density distribution
in Fig.~\ref{fig1} and Fig.~\ref{fig4}). A roughly flat region of
the curves where the bar size due to successive clipping
diminishes and becomes thinner, the $\Omega_{P}$ of both gas and
stellar bar is reaching $65$ km s$^{-1}$ kpc$^{-1}$. A rapid drop
in the curves corresponds to the clipping when the bar is not
continuous anymore, but rather looks like two bright patches,
after which the determination of the pattern speed is barely
possible. This behavior holds for several snapshots we have
analyzed between $2$ and $4$ Gyrs of evolution. The same
mechanism should be responsible for a rapid change in $\Omega_{P}$
for small $\Sigma_{\rm pert}$ values added to the intensity field.
Perturbations of the surface density of the the same amplitude as
of the faint disk and spiral arms lead to underestimation as shown
in Fig.~\ref{fig6}. Thus, we may conclude that the role of the
spiral arms in the application of the TW method is very important.
A similar conclusion was reached by \citet{Rand2004} who found
that too much clipping could violate the principal assumption of
the TW method and lead to incorrect pattern speed determinations.
On the other hand, we found that the restriction of the maximum
surface density (clipping ``from above'') produces the bifurcation
in the center of $\langle V\rangle$ vs. $\langle x\rangle$ plot
(not shown here), creating an effect of presence of two different
pattern speeds. The size of the bifurcated area increases with the
diminishing of the upper limit of the surface density. If these
points are omitted from the linear fit, the resulting pattern
speed is not affected.

The observations measure the intensity of H$\alpha$ emission
line rather than the gas surface density. The results above were
analyzed only for the surface density field of gas and stars
derived directly from the simulation, but not for the intensity
field produced by ionized gas. The original TW method is
formulated for the stellar component, where it is assumed that the
disk surface density is directly proportional to the number of
stars per unit area. However, in the case of the gas, the tracer
is sensitive to the square of the density because the regions of
ionized hydrogen will produce recombination lines whose rough
emissivity is given by
\begin{equation} I\sim \rho^2 T^{-1/2}
\label{eq2}
\end{equation}
If we assume that the gas is isothermal everywhere in the disk,
then the emission map will be a function of gas density squared
only. In this case, our simulations show that $\Omega_P$ is
overestimated by roughly $30$\%. Nonetheless, when the emission
map is clipped at the bar level (i.e., masking the spiral arms and
disk) the value of $\Omega_P$ is not changed. The same result was
found for the the observed H$\alpha$ emission map, presented in
the next section, if transformed into the surface density of the
gas. The clipping is important in this case because the
transformation given by equation \ref{eq2} mainly affects the
regions with high density gradients, i.e., transitions between bar
and disk.

These results establish the validity margins of the TW method for
our models. Although they were determined for a simple galaxy
model, they give us the general clues on parameter dependencies.
Complete hydrodynamical simulations, including star formation,
supernovae feedback and radiative cooling will be presented
elsewhere.

\section{Application to NGC 3367}

NGC 3367 is classified as a SBc type barred galaxy and it is
considered as an isolated galaxy at a distance of $43.6$ Mpc
($\mbox{H}_0=75$ km s$^{-1}$ Mpc$^{-1}$), located behind the Leo
Spur group of galaxies. This galaxy has a remote neighbor, NGC
3419, at a distance of $900\pm100$ kpc \citep{GarciaBarreto2001}.
At a radius of $10$ kpc from the nucleus there is an optical
structure consisting of several H$\alpha$ knots that resembles a
bow-shock \citep{GarciaBarreto1996a, GarcaBarreto1996b}.
\citet{GarcaBarreto1998} report Very Large Array (VLA)
observations with an angular resolution of $4.5\arcsec$ at $1.46$
GHz and find radio continuum emission from two lobes that extend
up to 6 kpc outside the plane of the disk, and a weaker emission
from the same disk. \citet{Gioia1990} also reported soft X-ray
emission from this galaxy.

\subsection{Ionized hydrogen kinematics data analysis}
We use the ionized hydrogen data cube from Fabry-Perot
interferometry observations published in
\citet{GarciaBarreto2001}. The data cube has dimensions
$512\times512\times48$, a final image scale of $0.58\arcsec$
pixel$^{-1}$, and spectral sampling of 19 km s$^{-1}$.
\citet{GarciaBarreto2001} used an interference filter with a
central wavelength of $6620$\AA, and a narrow band ($30$\AA) in
order to isolate the redshifted H$\alpha$ emission of this galaxy.
The authors made the calibration with a neon lamp centered at
$6598.95$\AA. The exposition time for each channel was of $120$ s.
No attempt was made by the authors for absolute calibration of the
H$\alpha$ emission.

\begin{figure*}[!htp]
\epsscale{2.3}\plottwo{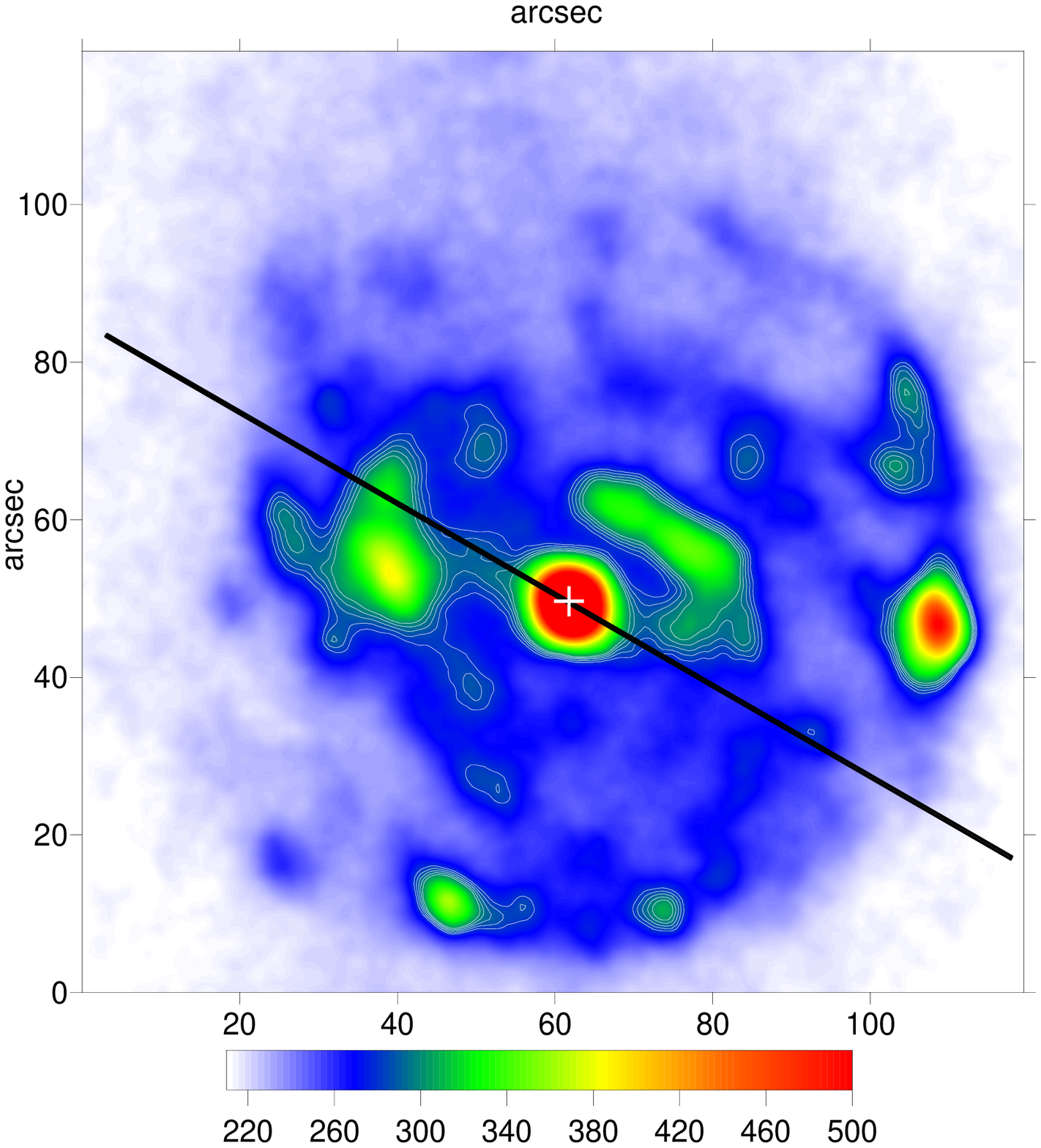}{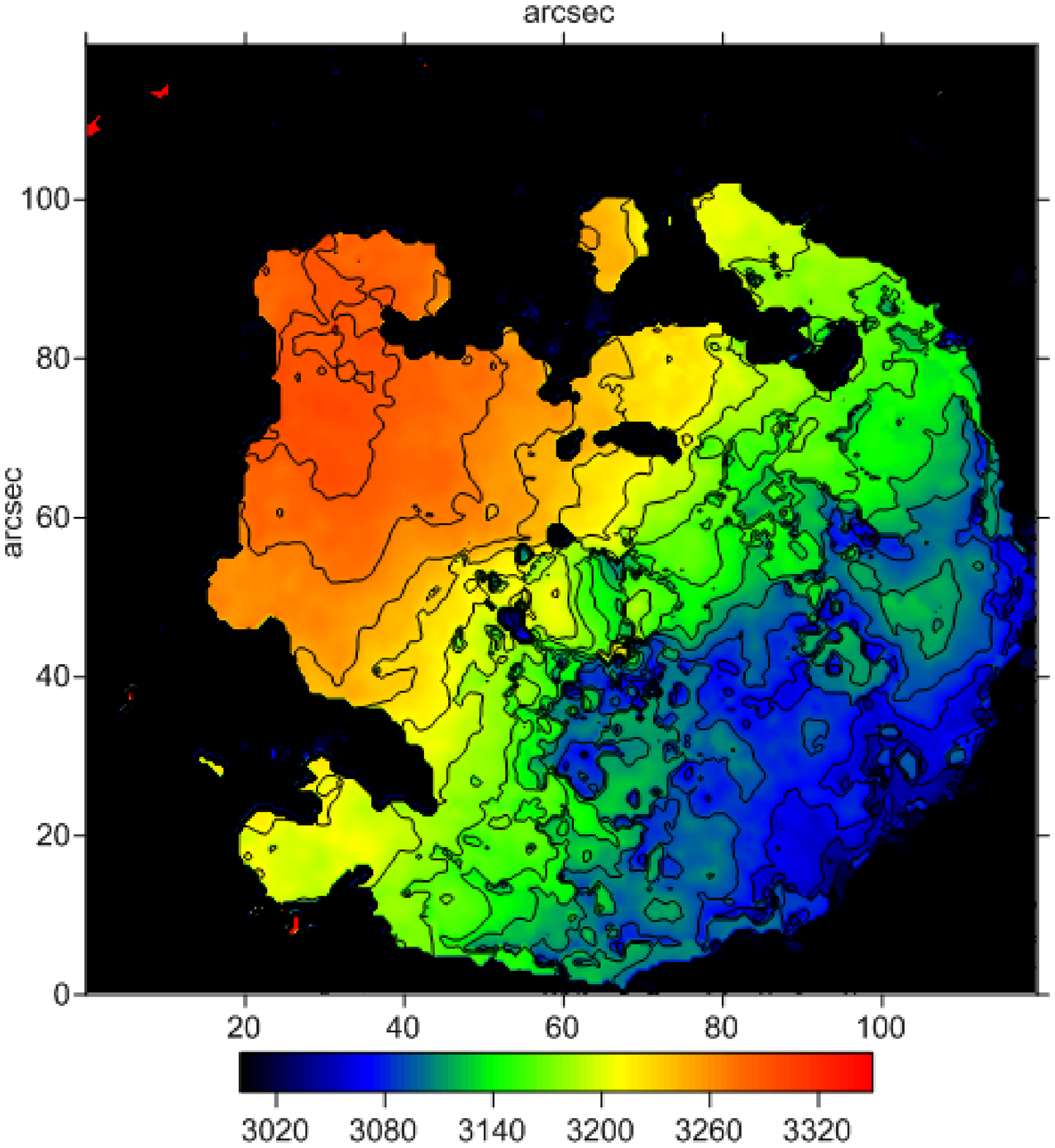}
%\epsscale{2.3}\plottwo{f8a.eps}{f8b.eps}
\figcaption{H$\alpha$ { image} (left) and radial velocity field of
NGC 3367 with contours levels separated by $20$ km s$^{-1}$
(right). The contours on the H$\alpha$ image depict the bar region
and are shown from $280$ to $300$ with step of $5$ in arbitrary
units. Also shown on the left panel is the position of the major
axis and the photometric center. The north direction is on top and
the east is on the left.\label{fig8}}
\end{figure*}

The data were already reduced with the
ADHOCw\footnote{http://www.oamp.fr/adhoc/adhocw.html developed by
J. Boulesteix.} software, and we use it to obtain the intensity
H$\alpha$ monochromatic map, the radial velocity field, and the
continuum map.  The H$\alpha$ image we use for our further
analysis was obtained by sum of $20$ channels, without
substraction of the continuum. The velocity field was obtained by
finding the barycenter of the H$\alpha$ profile peaks for each
pixel. The photometric center was determined from the continuum
map by looking for the brightest pixel, which in this case
coincides with the kinematic center \citep{FuentesCarrera2004,
FuentesCarrera2007}. We trimmed these maps to $204\times204$ to
exclude external sources and employ a spectral and spatial
smoothing with a Lorentzian function with a FWHM of $3$ pixels {
($\sim1.7\arcsec$ compared with the seeing $\sim 1.2\arcsec$)},
more adapted to the instrumental function of a Fabry-Perot. The
final H$\alpha$ image and the radial velocity field are shown in
(Fig. \ref{fig8}).

We use the two dimensional radial velocity field to obtain a
rotation curve for NGC 3367. \citet{GarciaBarreto2001} already
obtained a rotation curve for this galaxy using the AIPS package.
For this work we wish to explore the variation of the parameters
of the rotation curve in order to optimize them because they are
of fundamental importance for determining the bar pattern
velocity. For this purpose we use the ADHOCw software. These
parameters are: inclination $i$, position angle PA of the major
kinematic axis, systemic velocity $V_{syst}$ and kinematic center.
We obtain values similar to those published in
\citet{GarciaBarreto2001} with the difference that our systemic
velocity and position angle are slightly higher: $V_{syst}=3164\pm
10$ km s$^{-1}$ instead of $V_{syst}=3030\pm 8$ km s$^{-1}$,
$PA=60\degr\pm5\degr$ instead of $PA=51\degr\pm 3\degr$. Our
kinematic center is located at R.A.$=10^{h}46^{m}35^{s}$ and
Dec.$=13\degr 45\arcmin 00\arcsec$ (J$2000.0$), which are the same
values reported by \citet{GarciaBarreto2001}. The inclination
with respect to the plane of the sky is $i=30\degr\pm5\degr$ as
in \citet{GarciaBarreto2001}. As reported by
\citet{GarciaBarreto2007} the stellar bar length is
$32\arcsec$ ($6.7$ kpc) and is oriented $15^{\circ}\pm 5^{\circ}$
from the kinematic major axis.

\begin{figure*}[!htp]
\plottwo{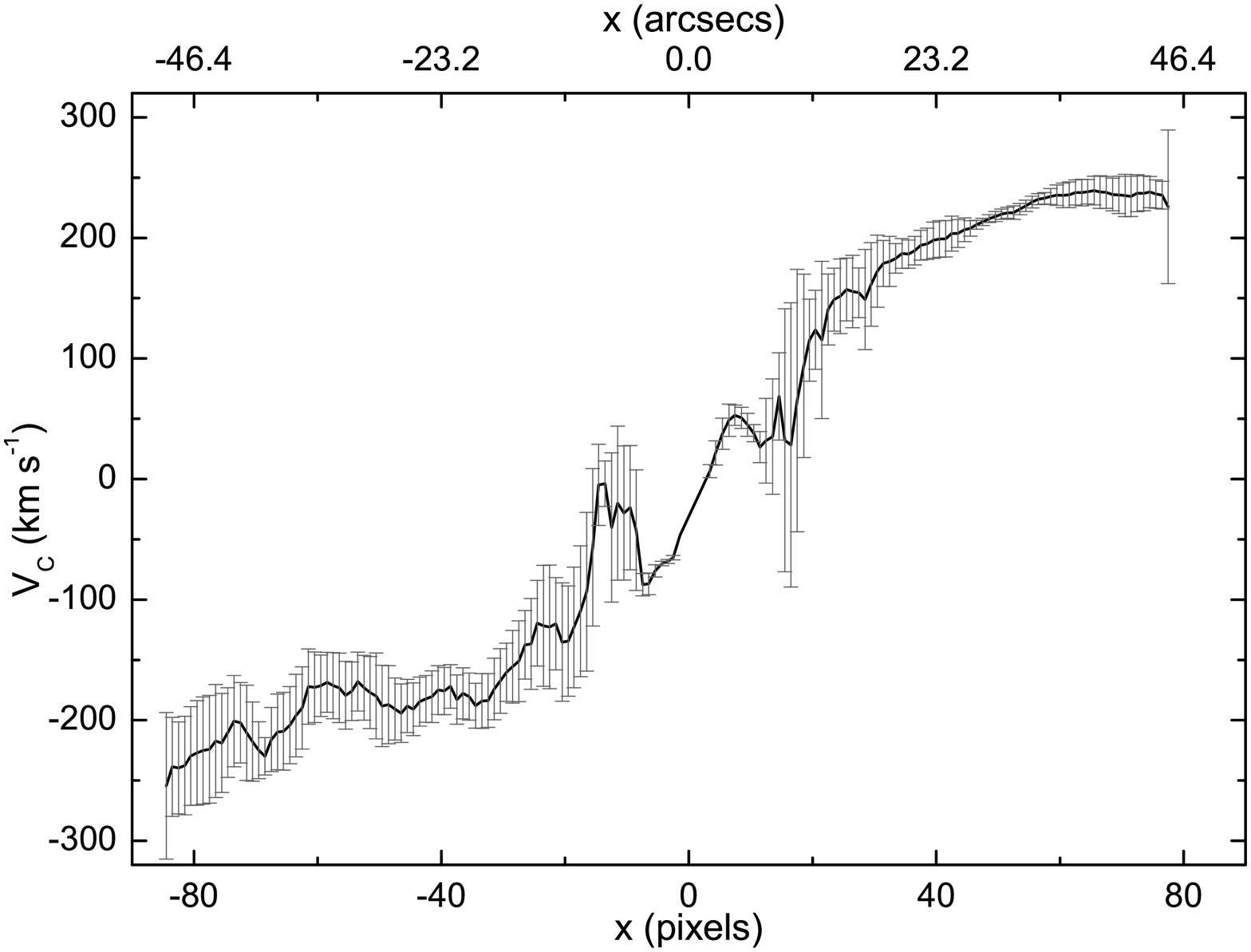}{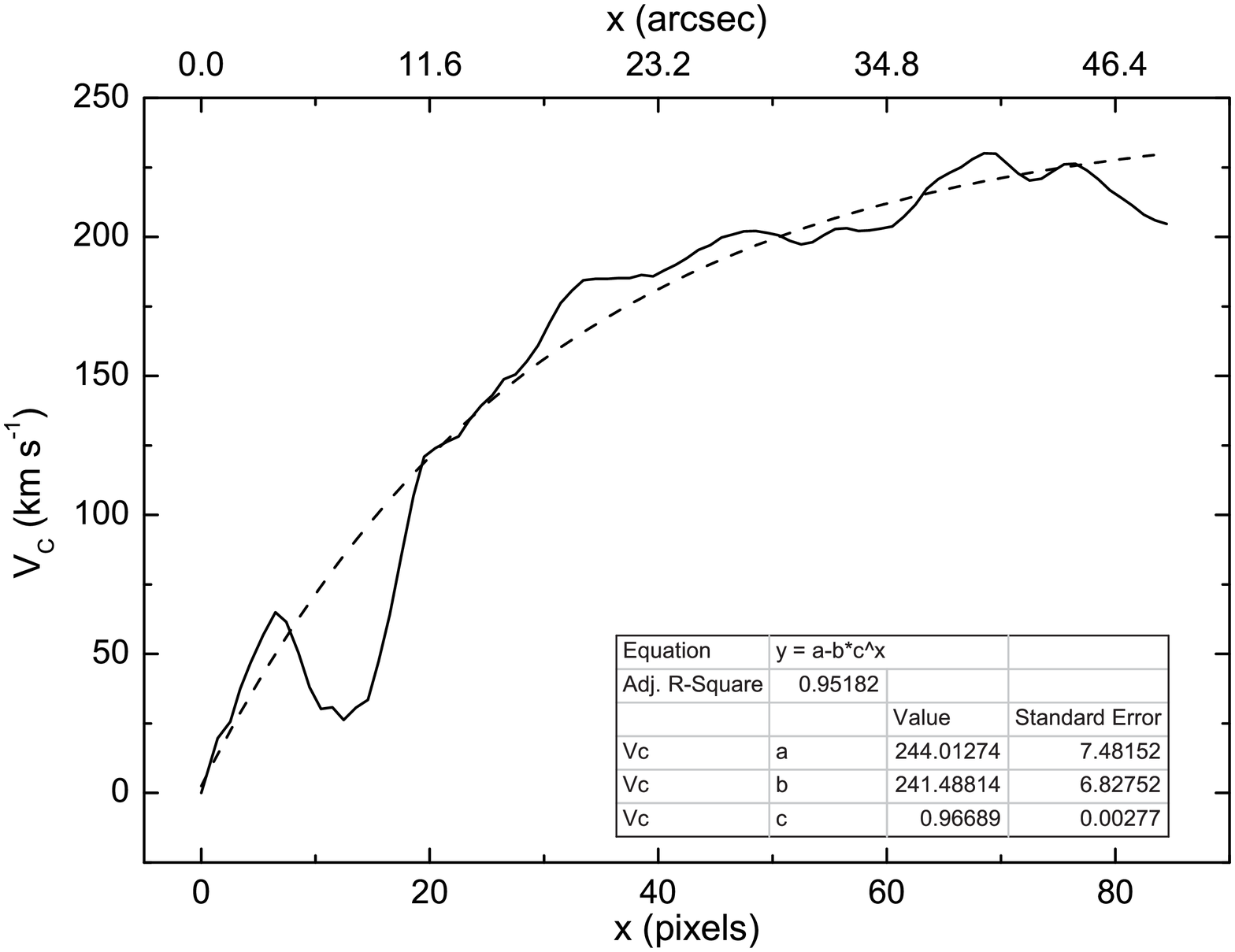} \figcaption{Full rotation
curve of NGC 3367 (left panel) and averaged rotation curve (right
panel, solid line) with overplotted fit (dashed line). See the
text for further details. \label{fig9}}
\end{figure*}

The rotation curve was determined by averaging velocities in two
sectors along the kinematic line of nodes. Once obtaining the full
rotation curve of NGC 3367, we average the rotation curve over
both sectors. We fit the mean rotation curve using a weighted
asymptotic regression model. The rotation curve of NGC 3367, the
averaged rotation curve and the fit are shown in Fig.~\ref{fig9}.

\section{Application of the TW method to NGC 3367}

We apply the Tremaine-Weinberg method to NGC 3367 in order to
measure the bar pattern speed. We built an IDL
program that calculates the intensity-weighted velocity
$\left<V\right>$ and the intensity-weighted position
$\left<x\right>$ for each strip along the kinematic minor axis of
the galaxy. Here the H$\alpha$ intensity serves as a weighting
function under the assumption that the surface density of the disk
is proportional to the H$\alpha$ intensity ($\Sigma_{disk} \propto
I_{H\alpha}$).

The TW method involves many parameters among which we consider the
position angle of the kinematic major axis (PA), the minimum and
maximum of the surface brightness image, position of the kinematic
center in pixels, systemic velocity, inclination, the range of
integration along the major axis and the location and number of
apertures along which we calculate the quantities $\left<V\right>$
and $\left<x\right>$. In order to explore the errors in the
$\Omega_{P}$ determination associated with the uncertainties in
the parameters we vary each of them to build a range of values. We
assume that within each determined interval there is a subset of
values where $\Omega_{P}$ is trustworthy if the plots of the
weighted mean velocity and position show little scatter such that
the slope in the $\left<V\right>$ vs. $\left<x\right>$ plot is
well fitted and passes though the origin. On the other hand, in
our case the curves of intensity-weighted velocity and position
should start from negative values and move smoothly towards
positive ones. By varying the full set of parameters we found that
four of them are actually relevant. These are the PA, the minimum
of the H$_{\alpha}$ image, the length of the slit along the
kinematical major axis and the range of variation along the minor
axis of the galaxy. The errors in $V_{syst}$, and kinematic center
have smaller effect on $\Omega_{P}$, as was already noted by
\citet{Merrifield1995}.

\begin{figure}[!htp]
\epsscale{1.0}\plotone{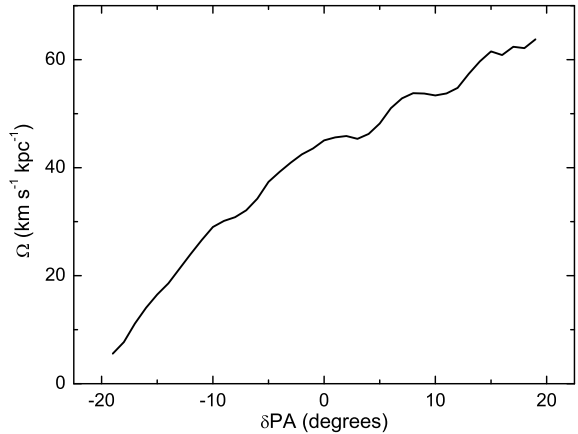} \figcaption{Variation of
$\Omega_{P}$ with position angle of the disk of NGC
3367.\label{fig10}}
\end{figure}

Thus, we begin with the variation of PA. We establish the origin
of the Cartesian coordinate system on the photometric center of
the H$_{\alpha}$ image of NGC 3367, such that the $x$ axis is
aligned with the disk kinematic major axis. First, we vary the PA
within the range of errors determined from the ADHOCw package
($\pm 5\degr$) in both north-east direction (positive) and in the
south-east direction (negative). From Fig.~\ref{fig10} it can be
noted that a variation of $\delta$PA by $\pm5\degr$ leads to
errors in $\Omega_P$ of $\pm 6$ km s$^{-1}$. We extend this range
up to $19\degr$ in order to compare it with the behavior of
$\Omega_P$ observed in the simulation. The result presented in
Fig.~\ref{fig10} shows that for $\delta$PA$>0$ the curve is almost
linear and roughly similar to the Fig.~\ref{fig2}. As the bar
becomes aligned with the kinematic major axis ($\delta$PA$<0$),
the $\Omega_P$ approaches zero faster. A similar conduct is found
for the simulated bar oriented in the same way as the observed one
($\sim15\degr$ from the major axis).

\begin{figure}[!htp]
\epsscale{1.0}\plotone{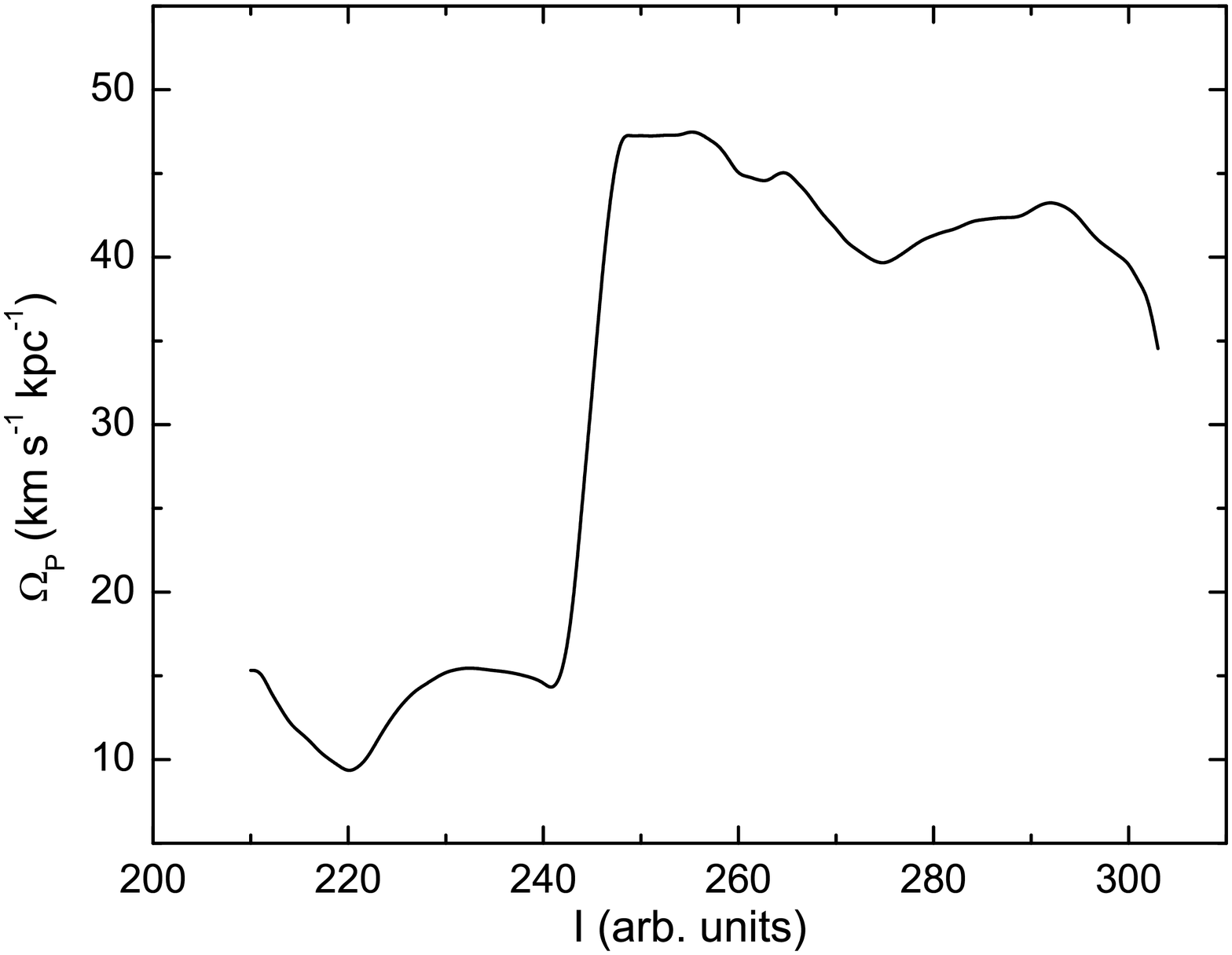} \figcaption{Variation of
$\Omega_{P}$ with the minimum of intensity of the H$_{\alpha}$
image. \label{fig11}}
\end{figure}

In order to select only those emission regions near the bar, we
mask the H$_{\alpha}$ image by changing the minimum of the image.
Clipping to a given minimum H$_{\alpha}$ intensity is necessary
because if the entire image is taken the plot of $\left<V\right>$
vs. $\left<x\right>$ is too noisy and it is difficult to establish
a clear fit. As shown in the previous section a sufficiently noisy
background could strongly affect the results. The masking
procedure helps to hide bright knots within the interarm regions
and allows to achieve higher signal to noise ratio. Additionally,
as demonstrated by \citet{Rand2004}, intensity clipping helps to
improve the results by removing the scatter produced by clumps,
but at the same time it could also remove part of the bar pattern.
Fig.~\ref{fig11} shows the variation of $\Omega_{P}$ with the
minimum of the H$_{\alpha}$ image. The range of varied intensity
corresponds to emission that roughly traces disk, spiral arms,
and the bar region, except the bright bulge. When compared with
the Fig.~\ref{fig7} a similar behavior of the pattern speed due to
clipping can be observed. Within the bar region (excluding disk
and arms) the errors due to clipping are $\pm5\%$. The small
variation of $\Omega_{P}$ within the clipping range $210-240$ is
probably due to the clumpiness of the H$_{\alpha}$ image around
the bar. We also examined the clipping of the maximum intensity
and found that it did not significantly alter the results.

Strong emission along the spiral arms represent a great source of
noise in bar pattern speed determination that one needs to isolate
in order to correctly perform the weighting procedure. The
emission intensity of spiral arms is much stronger than that of
the bar, and $\langle x\rangle$ and $\langle V\rangle$ will be
biased by the arms. For this reason we resort to constraining the
integration area uniquely to the bar zone. We restrict the range
of integration along the kinematical major axis to avoid the
region of the eastern arm. This is an extended and prominent zone
that biases significantly the signal to noise ratio of the
intensity near the north-east side of the bar (Fig.~\ref{fig8}).
Since we are interested in the bar pattern speed, avoiding spiral
arms that may have a pattern speed distinct from that of the bar
would also improve the signal. The fourth important parameter is
the range along the minor axis of the galaxy, i.e., the number of
apertures. If this range is not limited, the bright spiral arms of
the galaxy, in particular the one located at north-west, enter in
the computation and the bar pattern speed cannot be determined.

\begin{figure}[!htp]
\epsscale{1.0}\plotone{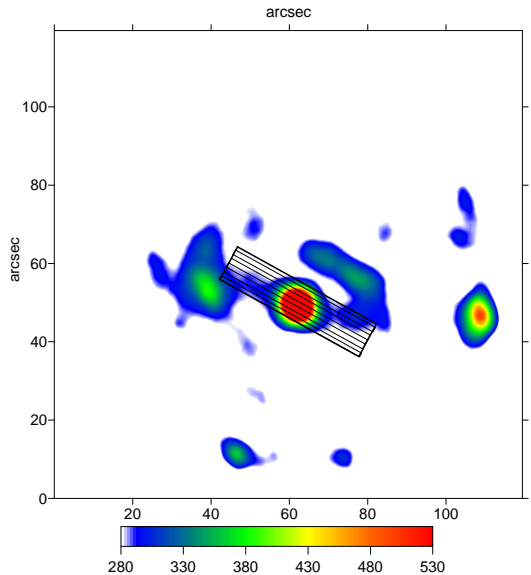}
%\epsscale{1.0}\plotone{f12.eps}
\figcaption{Masked H$_{\alpha}$ image and the area of integration
used for the final $\Omega_P$ calculation. Some of the apertures
used in the TW method were also marked. { Box units are in
arcseconds, and intensity is in arbitrary units}. \label{fig12}}
\end{figure}

After analyzing the parameter variation we determine the bar
pattern speed of NGC 3367. The parameters mentioned above were
constrained in the following sequence: limiting the integration
area, limiting the intensity, and varying the PA. The final masked
H$_{\alpha}$ image and the region where the $\Omega_P$ was
determined is shown in Fig.~\ref{fig12}. The averaging of
quantities in equation (\ref{eq1}) is done along strips of one
pixel width, totalling $18$ slits parallel to the kinematic major
axis within the box.

We find for NGC 3367 a value of $\Omega_P=43\pm6$ km s$^{-1}$ kpc$^{-1}$.
This value is consistent with the value reported in \citet{GarciaBarreto2001}.
\begin{figure*}[!htp]
\epsscale{2.0}\plottwo{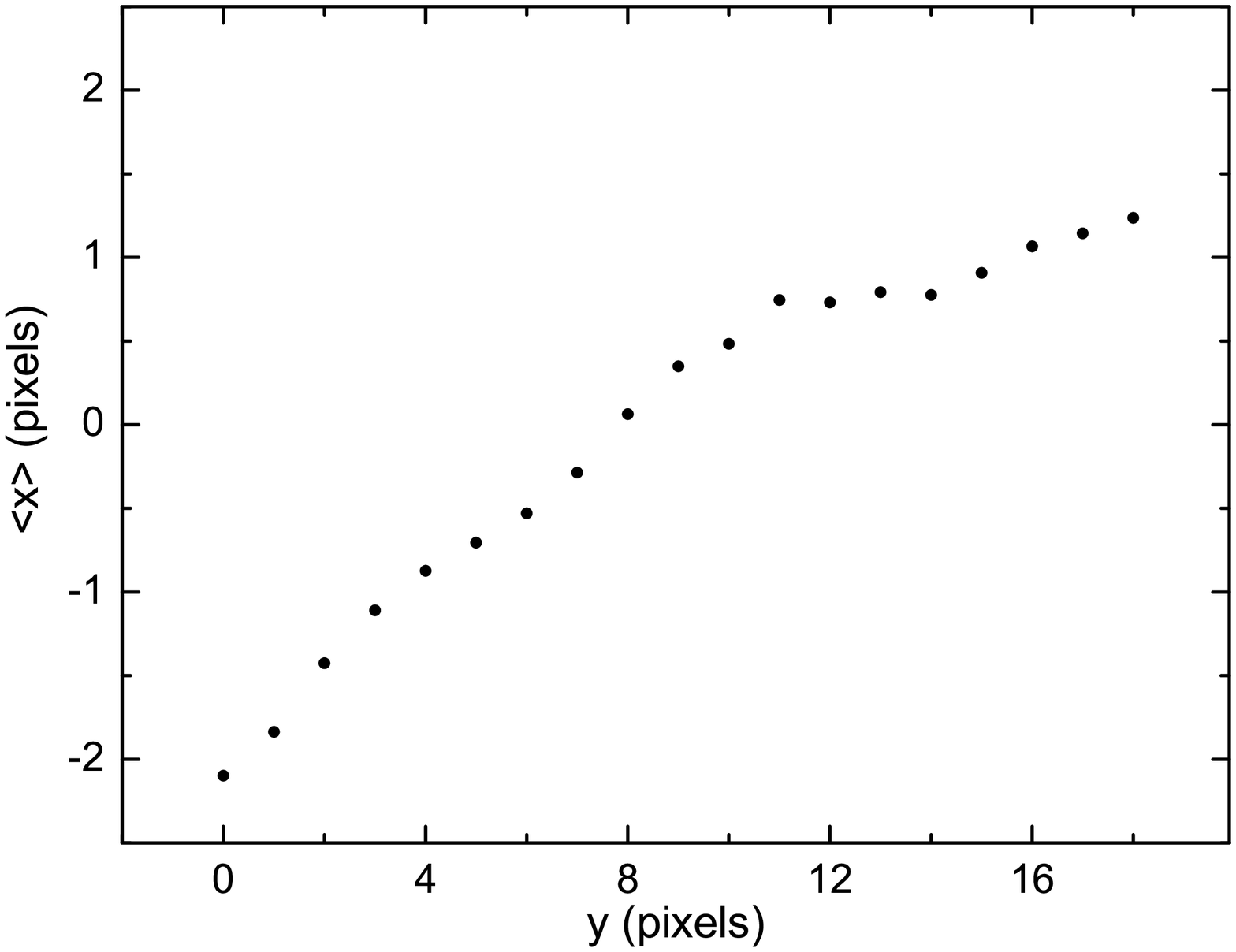}{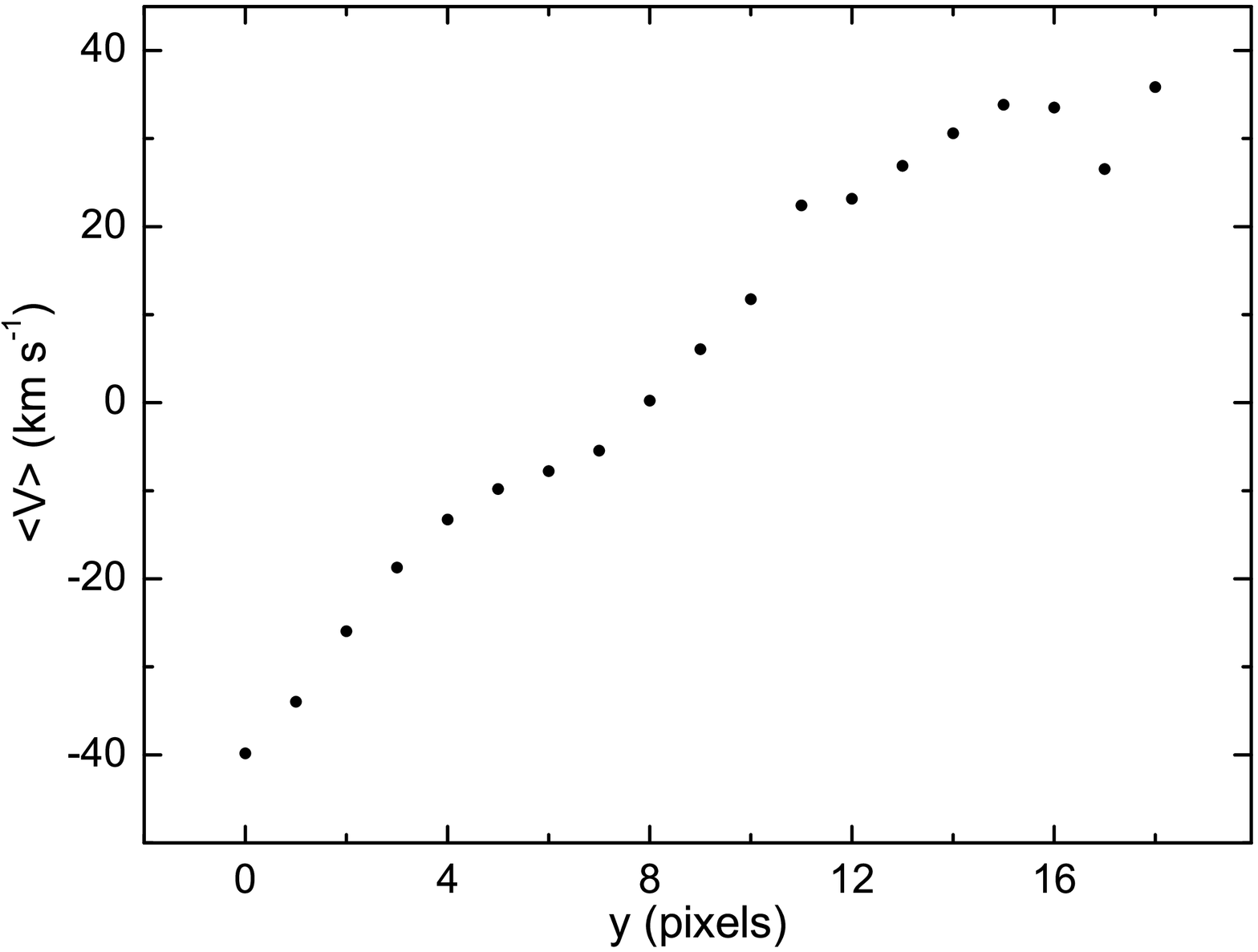} \figcaption{Weighted
average position (left) and weighted average velocity (right)
along the minor axis. \label{fig13}}
\end{figure*}
\begin{figure*}[!htp]
\epsscale{2.0}\plottwo{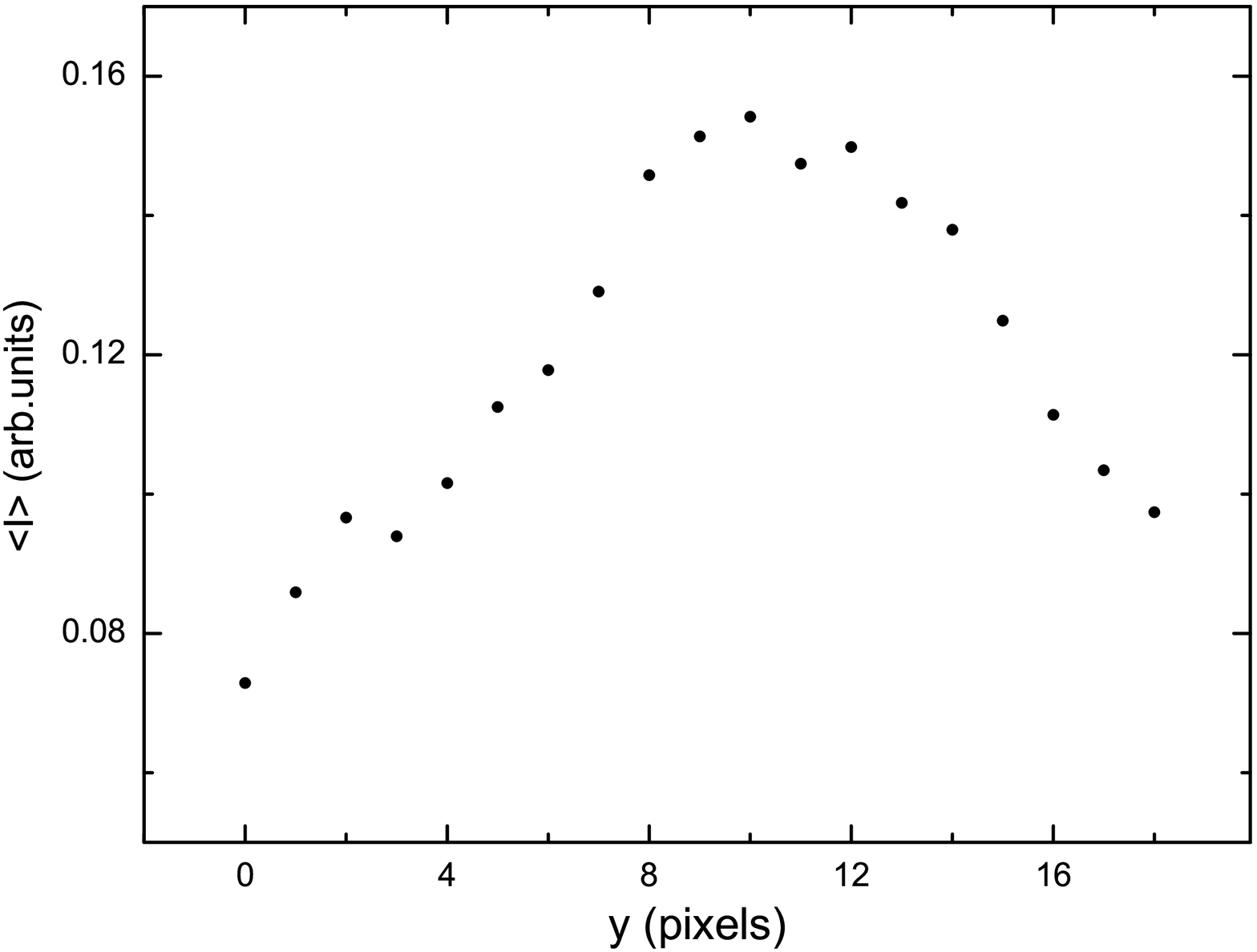}{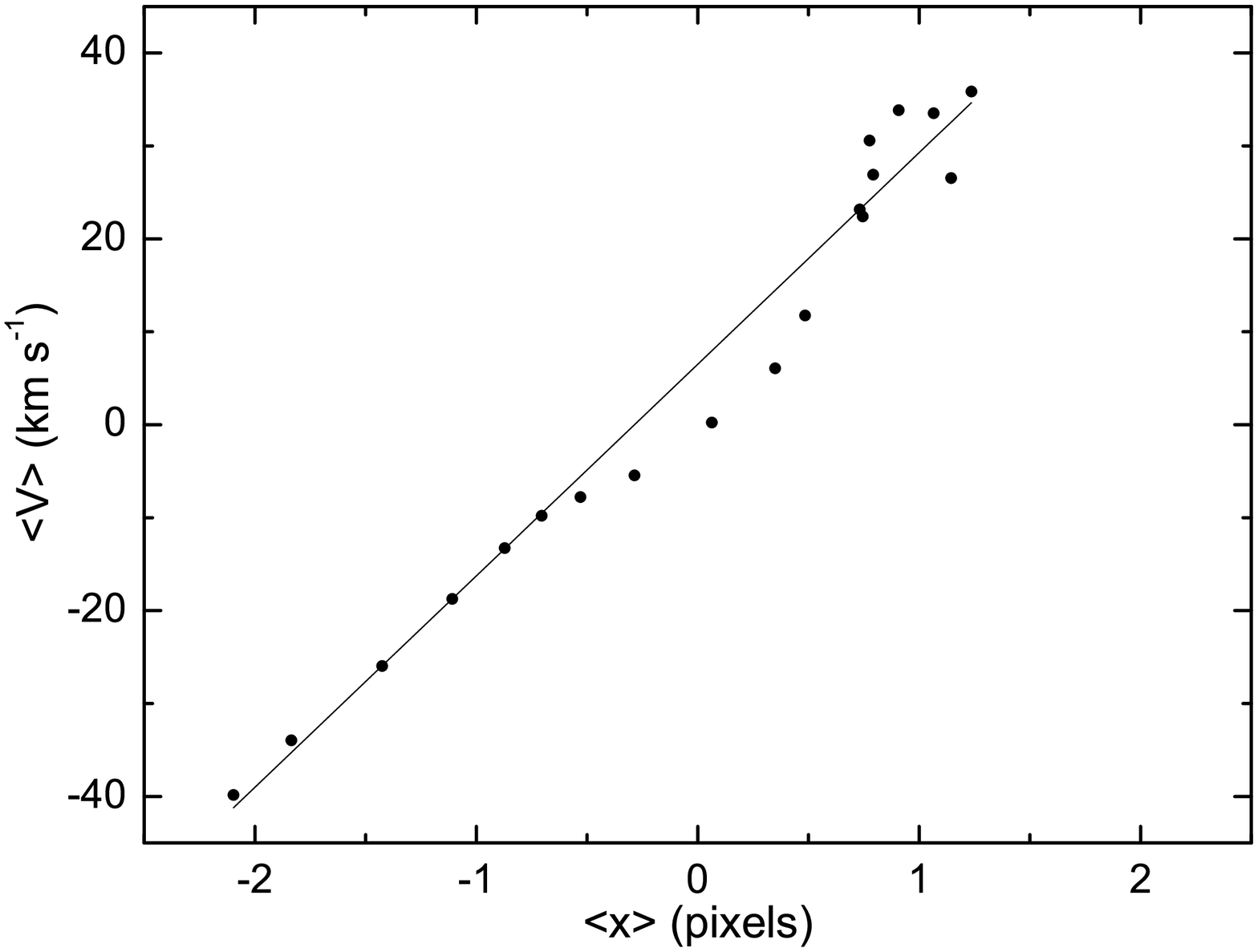} \figcaption{Mean
intensity along the slits (left) and weighted average velocity as
a function of the weighted average position (right).\label{fig14}}
\end{figure*}
In Fig.~\ref{fig13} we show the mean weighted position and the
mean weighted velocity as a function of the position along the
minor axis of the galaxy. Both curves show an almost linear trend,
except that $\left<x\right>$ shows flattening after $y=11$ pixels
($1.35$ kpc). This is probably due to a characteristic shape of
the bar, both parts of which are not symmetric and which appears
to be ``broken''. This asymmetry is also observed in broadband
ultraviolet images \citep{GarciaBarreto2007}. In Fig.~\ref{fig14}
we show the mean weighted velocity as a function of the mean
weighted position and the mean intensity as a function of the
position along the minor axis. The mean intensity has a bell shape
similar to what is observed for a simulated bar. The sequencing of
points also checks that we really measure the bar pattern speed.
It is worth noting that if a linear fit is applied to the plots in
Fig.~\ref{fig13} and then the slope is estimated, the pattern
speed is $\Omega_{P}\approx47$ km s$^{-1}$ kpc$^{-1}$.

\subsection{Determination of $\Omega_{P}$ by resonances analysis.}
\begin{figure}[!htp]
\epsscale{1.0}\plotone{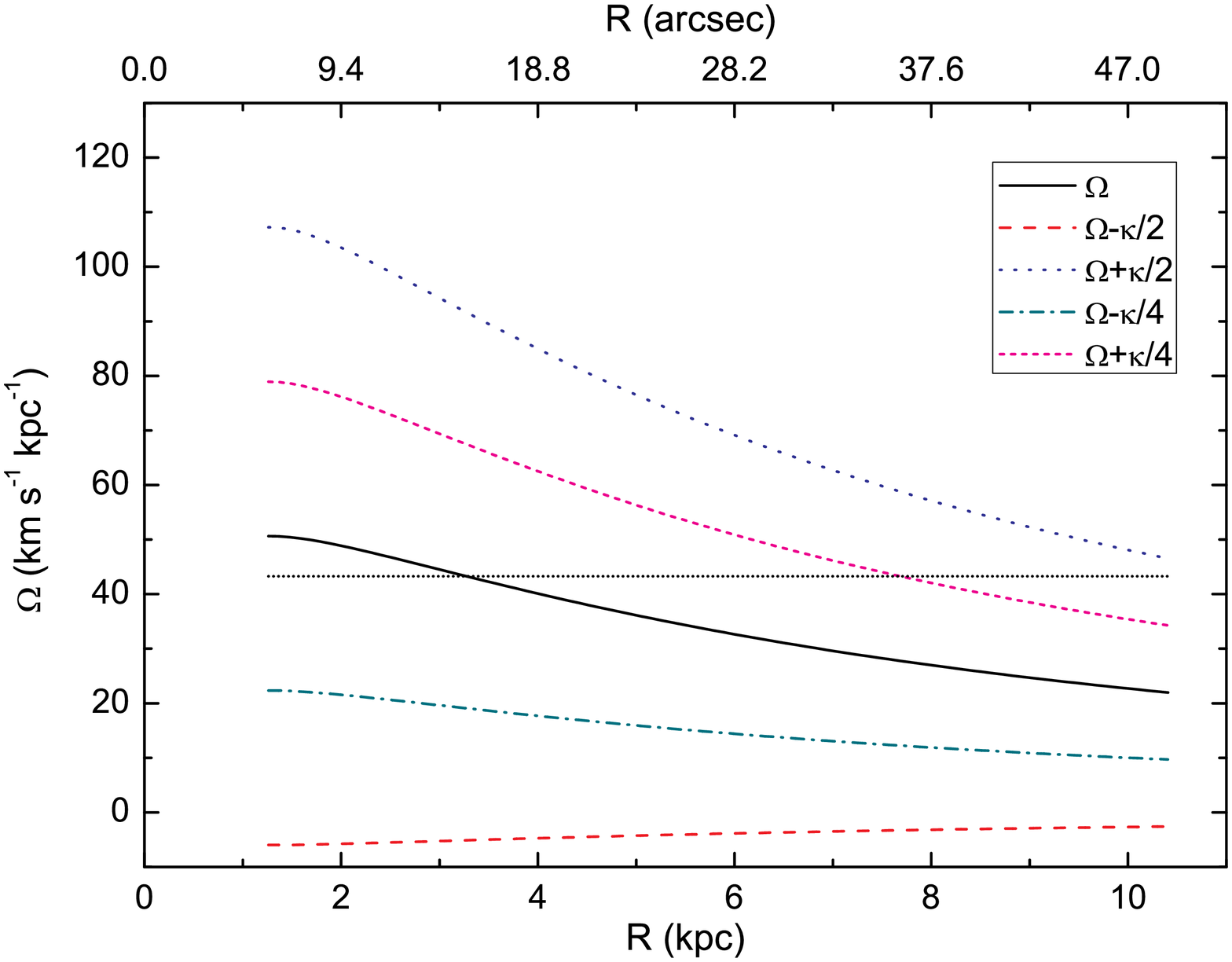} \figcaption{Resonances of NGC
3367. The dash-dotted horizontal line corresponds to the bar
pattern speed, $\Omega_P$, determined from the TW
method.\label{fig15}}
\end{figure}

We compared $\Omega_{P}$ obtained from the TW method with the one
obtained from analyzing fundamental resonances of NGC 3367
predicted by the linear theory. According to the linear theory at
the resonance radii there are expected rings of matter with
enhanced star formation caused by gas shocks. In particular, NGC
3367 has a ring of radius $\sim 7$ kpc formed by spiral arms
wrapped by more than $180\degr$. Using the diameter of the stellar
bar $6.7$ kpc given by \citet{GarciaBarreto2007} and assumption
that the bar ends near the corotation, we find that $\Omega_{P}=45
\pm7$ km s$^{-1}$ kpc$^{-1}$. Here, the errors in $\Omega_{P}$ are
due to the rotation curve determination and fitting procedure.
This value agrees within the error limits with the value we found
by the TW method and also accords with the value reported in
\citet{GarciaBarreto2001} of $\Omega_{P}=43$ km s$^{-1}$
kpc$^{-1}$. In Fig.~\ref{fig15} are shown the resonances for NGC
3367 and the bar pattern speed determined by the TW method. The
resonances were obtained from the fit to the rotation curve
(Fig.~\ref{fig9}) given in this work. According to the figure the
outer 4:1 resonance is located at $7.7$ kpc ($36.7\arcsec$), the
OLR is at some $11$ kpc ($52\arcsec$), and there is no ILR.
However, it should be noted that the resulting $\Omega_{P}$
derived in this way should be taken with caution because the ring
is not necessarily located at corotation. On the other hand, given
the complex structure of the rotation curve in the center and 
limited resolution, it is difficult to judge whether an ILR is
absent or present.

\section{Discussion}
The main aim of this work is to estimate the bar pattern speed of
NGC 3367 galaxy using H$\alpha$ kinematic data. For this purpose
we first investigate the sensitivity of results to the data
quality on numerical galaxy models. We have not intended to create
an exact numerical model of NGC 3367, but rather to use the
simulations as toy models. For this reason we cannot directly
compare the lengths and pattern speeds of simulated and observed
bars. We have tested the TW method for a simulated galaxy bar
against contamination of surface density and velocity fields by
white noise and Gaussian random field perturbations in our
simulated data. We found that the surface density is critical to
noise and the bar area should have a good signal to noise ratio in
order to obtain reliable results, because it is used as a
weighting function in equation (\ref{eq1}). In contrast, the
velocity field appears to be less sensitive to errors, although
they also affect the accuracy of the results. This is due to the
zero net velocity perturbation introduced in calculus of
$\left<V\right>$ where the errors are partially cancelled out.
However, for gas-poor, old stellar bars, where the buckling is
significant, $v_z$ component of the velocity field may be
important. We also checked the influence of the position angle on
the resulting $\Omega_P$. In agreement with \citet{Debattista2003}
the results are sensitive to PA uncertainties. When the disk major
axis and the bar become aligned, the errors in $\Omega_P$ increase
faster than in the opposite case.

Based on the results of the robustness test, we apply the TW
method to NGC 3367 using published 2D data of Fabry-Perot
interferometry. After carefully studying the parameters that could
affect the results (minimum intensity, range of integration,
number of strips, high intensity knots and arms, position of
kinematic major axis), we found $\Omega_P=43\pm6$ km s$^{-1}$
kpc$^{-1}$. We also determine the bar pattern speed by means of
resonances analysis and found $\Omega_{P}=45 \pm7$ km s$^{-1}$
kpc$^{-1}$. These two results are consistent and in agreement with
the value reported previously by \citet{GarciaBarreto2001}. The
trend in variation of $\Omega_{P}$ with the position angle of the
kinematical major axis is consistent with the result encountered
in the simulated barred galaxy.

As shown in Sect.~\ref{twtest} the variation of the minimum of the
surface density for the simulated barred galaxy could
significantly affect $\Omega_{P}$ determination. As we have
demonstrated, for NGC 3367 the adopted minimum of intensity in
H$_{\alpha}$ image is one of the fundamental parameters. Applying
a clipping allows us to increase the contrast of emission in the
interarm region, thus improving the signal associated with the
bar. But, on the other hand, if the trend found for the simulated
bar applies for H${\alpha}$ gaseous bar, the clipping we used for
NGC 3367 could lead to an overestimation of the $\Omega_{P}$ by
more than $50\%$ and the bar would lie within a flat region of
the curves shown in Fig.~\ref{fig7}. Thus, if the clipping is not
applied, our value of $\Omega_{P}$ indeed can be as small as $\sim
15$ km s$^{-1}$. Yet, a deeper study using galaxy models with
complex gas physics is required to verify whether the pattern
speed is underestimated due to clipping. The other two parameters
we investigated essentially restrain the range of integration
parallel and perpendicular to the kinematic major axis. These are
underlying parameters that allow us to significantly improve the
signal to noise ratio near the bar zone.

When observing the radial velocity field (Fig.~\ref{fig8}), an
interesting feature can be noted in the nuclear region. The
isovelocity contours in the center of the galaxy that are almost
perpendicular to the main ones may imply non-circular motions.
Since the sharp turn in isophotes is frequently associated with a
secondary bar, the possibility of non-circular motions due to
different origins is present. In particular, the inner gas polar
ring or disk, or a secondary bar may be responsible for the
characteristic shape of the isovelocity contours. Such behavior
has been observed previously in several galaxies, see for example
\citet{Moiseev2004}. However,  due to lack of resolution, in
this work we were unable to characterize and identify any
secondary pattern speed. A study with a better angular resolution
and a comparison with the stellar counterpart is necessary in
order to analyze the detailed kinematical features in the center
of this galaxy.

\section{Conclusions}
In this work we apply the TW method to a simulated barred galaxy
and test the sensitivity of the method to possible sources of
error. Then we measure the bar pattern speed of NGC 3367 using the
same method. The results obtained from the application of this
method to a simulated galaxy in general hold also for the observed
one. However, the simulations have shown that a significant error
can emerge due to applying the clipping to the surface density.
These facts motivated us to assert some important conclusions
within the limits of our work. First of all, bearing in mind the
principal assumptions of the TW method, we can say that it works
well enough for the gas phase whenever it is continuously
distributed along the bar. On the other hand, the validity of
applying clipping to the intensity map should be further
investigated. We also want to note the importance of two
dimensional data that allow the exploration of a wider range in
parameters variation. Given the errors of the position angle
determination the variation of $\delta$PA$\sim 5\degr$ do not
affect too much the resulting pattern speed for NGC 3367.
Finally, determining the bar pattern speed by locating the
corotation resonance, we found that the result is similar to that
obtained by application of the TW method. This fact supports in
part the reliability of the method in the case of ionized gas
data. A deeper study of this galaxy would be worth, in particular,
the comparison with the pattern speed determined from the stellar
long-slit spectral observations, which would verify the validity
of the TW method for H$\alpha$ kinematic data.

\acknowledgments R.Gabbasov is supported by postdoctoral
fellowship provided by UNAM. P.Repetto acknowledges CONACyT for
doctoral scholarship. M.Rosado acknowledges CONACyT, project
number 46054-F, and DGAPA, project number IN 100606. The
simulations were performed on KanBalam computer at DGSCA,
departament of supercomputing of UNAM.

\nocite{*}
\bibliographystyle{apj}
\bibliography{bib}
\end{document}